\documentstyle[12pt]{article}
\input{epsf}

\newcommand{\xxxcA}{{\cal A}}
\newcommand{\wil}{W(\sigma)}

\newcommand{\half}{\frac{1}{2}}
\newcommand{\xxxcal}[1]{{\cal#1}}

\newcommand{\sect}[1]{\setcounter{equation}{0}\section{#1}}

\def\rf#1{(\ref{eq:#1})}
\def\lab#1{\label{eq:#1}}
\def\nonu{\nonumber}
\def\br{\begin{eqnarray}}
\def\er{\end{eqnarray}}
\def\be{\begin{equation}}
\def\ee{\end{equation}}

\def\lb{\lbrack}
\def\rb{\rbrack}

\def\({\left(}
\def\){\right)}

\newcommand{\ct}[1]{\cite{#1}}
\newcommand{\bi}[1]{\bibitem{#1}}

\relax

\def\Tr{\mathop{\rm Tr}}

\newcommand{\sbr}[2]{\left\lbrack\,{#1}\, ,\,{#2}\,\right\rbrack}

\def\a{\alpha}

\def\b{\beta}
\def\bfphi{{\bf \phi}}
\def\by{{\bar y}}
\def\bz{{\bar z}}

\def\d{\delta}

\def\G{\Gamma}

\def\h{{1\over 2}}
\def\l{\lambda}

\def\o{\over}

\def\O{\Omega}

\def\pa{\partial}
\def\pr{\prime}

\def\ra{\rightarrow}
\def\s{\sigma}
\def\S{\Sigma}

\def\thh{{\tilde H}}

\def\tp0{\Theta_{+}^{(0)}}
\def\tm0{\Theta_{-}^{(0)}}

\def\u2{\mid u\mid^2}

\def\vp{\varphi}

\def\ca{{\cal A}}

\def\cd{{\cal D}}

\def\cg{{\cal G}}

\def\cs{{\cal S}}
\def\cw{{\cal W}}
\def\ctt{{\cal T}}

\def\flie{{\cal F}}

\def\f#1#2#3 {f^{#1#2}_{#3}}

\def\win1{{\sf w_{1+\infty}}}

\def\Win1{{\sf W_{1+\infty}}}

\def\rlx{\relax\leavevmode}
\def\inbar{\vrule height1.5ex width.4pt depth0pt}
\def\IZ{\rlx\hbox{\sf Z\kern-.4em Z}}
\def\IR{\rlx\hbox{\rm I\kern-.18em R}}
\def\IC{\rlx\hbox{\,$\inbar\kern-.3em{\rm C}$}}
\def\IN{\rlx\hbox{\rm I\kern-.18em N}}
\def\IO{\rlx\hbox{\,$\inbar\kern-.3em{\rm O}$}}
\def\IP{\rlx\hbox{\rm I\kern-.18em P}}
\def\IQ{\rlx\hbox{\,$\inbar\kern-.3em{\rm Q}$}}
\def\IF{\rlx\hbox{\rm I\kern-.18em F}}
\def\IG{\rlx\hbox{\,$\inbar\kern-.3em{\rm G}$}}
\def\IH{\rlx\hbox{\rm I\kern-.18em H}}
\def\II{\rlx\hbox{\rm I\kern-.18em I}}
\def\IK{\rlx\hbox{\rm I\kern-.18em K}}
\def\IL{\rlx\hbox{\rm I\kern-.18em L}}
\def\one{\hbox{{1}\kern-.25em\hbox{l}}}
\def\0#1{\relax\ifmmode\mathaccent''7017{#1}%
B        \else\accent23#1\relax\fi}

\def\PRL#1#2#3{{\sl Phys. Rev. Lett.} {\bf#1} (#2) #3}
\def\NPB#1#2#3{{\sl Nucl. Phys.} {\bf B#1} (#2) #3}

\def\CMP#1#2#3{{\sl Commun. Math. Phys.} {\bf #1} (#2) #3}
\def\PRD#1#2#3{{\sl Phys. Rev.} {\bf D#1} (#2) #3}

\def\PLB#1#2#3{{\sl Phys. Lett.} {\bf #1B} (#2) #3}
\def\JMP#1#2#3{{\sl J. Math. Phys.} {\bf #1} (#2) #3}

\def\PR#1#2#3{{\sl Phys. Reports} {\bf #1} (#2) #3}

\def\FAaIA#1#2#3{{\sl Functional Analysis and Its Application} {\bf #1} (#2)
#3}

\def\TMP#1#2#3{{\sl Theor. Mat. Phys.} {\bf #1} (#2) #3}

\def\PHSD#1#2#3{{\sl Physica} {\bf D#1} (#2) #3}
\def\PJA#1#2#3{{\sl Proc. Japan. Acad} {\bf #1A} (#2) #3}

\begin{document}
\begin{titlepage}
\vspace*{-1cm}
\noindent
October 1997 \hfill{US-FT/31-97}\\
\phantom{bla}
\hfill{IFT-P/066/97} \\
\phantom{bla}
\hfill{UMTG-202}\\
\phantom{bla}
\hfill{\tt hep-th/9710147}

\vspace{.1in}
\begin{center}
{\large\bf A New Approach to Integrable Theories in any Dimension}
\end{center}

\begin{center}
Orlando Alvarez$^{\ast}$,
Luiz A. Ferreira$^{\dagger}$
 and
J. S\'anchez Guill\'en$^{\sharp}$

\vspace{.3 cm}
\small
\par \vskip .1in \noindent
$^{\ast}$Department of Physics\\
University of Miami\\
P.O. Box 248046\\
Coral Gables, FL 33124, USA

\par \vskip .1in \noindent
$^{\dagger}$Instituto de F\'\i sica Te\'orica - IFT/UNESP\\
Rua Pamplona 145\\
01405-900  S\~ao Paulo-SP, BRAZIL

\par \vskip .1in \noindent
$^{\sharp}$Departamento de F\'\i sica de Part\'\i culas,\\
Facultad de F\'\i sica\\
Universidad de Santiago\\
E-15706 Santiago de Compostela, SPAIN
\normalsize
\end{center}
\vspace{.2in}
\begin{abstract}
The zero curvature representation for two dimensional integrable 
models is generalized to spacetimes of dimension $d+1$ by the 
introduction of a $d$-form connection.  The new generalized zero 
curvature conditions can be used to represent the equations of motion 
of some relativistic invariant field theories of physical interest in 
$2+1$ dimensions (BF theories, Chern-Simons, $2+1$ gravity and the 
$CP^1$ model) and $3+1$ dimensions (self-dual Yang-Mills theory and 
the Bogomolny equations).  Our approach leads to new methods of 
constructing conserved currents and solutions.  In a submodel of the 
$2+1$ dimensional $CP^1$ model, we explicitly construct an infinite 
number of previously unknown nontrivial conserved currents.  For each 
positive integer spin representation of $sl(2)$ we construct $2j+1$ 
conserved currents leading to $2j+1$ Lorentz scalar charges.
\end{abstract}
\end{titlepage}

\sect{Introduction}
\label{sec:intro}

We propose a generalization to higher dimensional relativistic field 
theories of the zero curvature condition used to solve integrable 
models in $(1+1)$ dimensions, based on a geometrical principle 
formulated in loop space.  The approach has many applications beyond 
that systematic formulation of known theories.  It provides a 
criterion for $d>2$ integrability, which yields new integrable models 
and it can be used to construct an infinite number of nontrivial 
conserved currents and new solutions as we show for some physically 
interesting cases.

The equations 
of motion for many models in two dimensions such as the sine-Gordon 
model were formulated in terms of the zero curvature condition
\begin{equation}
[\partial_0 + A_0 , \partial_1 + A_1 ]=0 \;.
\lab{2dint}
\end{equation}
This formulation first appeared  in the work on the nonlinear 
Schr\"odinger equation (NLS) by Zakharov and Shabat \ct{zs} in 1971 and
opened the way to obtaining many results on the NLS system including 
solutions and conserved charges.

Soliton theory is now a mature area of research and we have a very 
good understanding of classical soliton solutions.  A common feature 
of multi-soliton solutions in most of soliton theories is that they 
are associated to vertex operator representations of affine Lie 
algebras and they correspond to some special points on the orbit, 
under the dressing transformation group, of some vacuum solutions 
\ct{BB,kac,jimbo,nos}.  Most of these developments have used in an 
essential way the zero curvature condition \rf{2dint}, but 
unfortunately these remarkable developments have been only achieved 
in one or two dimensional theories.  This formulation appears not to 
generalize directly to dimensions higher than two, although there are 
some exceptional cases, like self-dual Yang-Mills \ct{BZAK} and a few 
nonrelativistic systems \ct{Novikov}.  One may ask the reasons why it 
has been difficult to extend these ideas to higher dimensions.  
Mathematicians and physicists may give different but in general 
complementary arguments.  From the point of view of high energy 
physicists there is one quite compelling argument.  Physicists are 
interested in Lorentz invariant local field theories where one has 
general results such as the TCP theorem and the Coleman-Mandula 
theorem which impose strong constraints in dimensions larger than two.  
Lorentz invariant local theories are subject to tough restrictions on 
the interplay between internal and space-time symmetries, and on the 
consequences of higher spin fields and higher spin conserved charges.  
For the integrable two dimensional models, the infinite number of 
conserved quantities may be related to highly nonlinear (and possibly 
nonlocal) symmetry transformations which very likely mix in a 
nontrivial way internal and space-time symmetries.  In general such 
transformations are not known.   
Direct methods for constructing conserved charges were developed which 
are independent even on the existence of the local symmetry 
transformations.  In this sense Noether's theorem is of little help in 
nonlinear two dimensional integrable models.  Additionally, the 
conserved charges of such models are in general high order polynomials 
in the momenta and derivatives of the fields.  Their Poisson bracket 
algebra is generally described by a type of $W$-algebra \ct{walg} 
which formally is not even an algebra.  Therefore one does not expect 
all these structures to appear in theories of physical interest in 
dimensions higher than two.

In this paper we show that some of the methods and results  used in two
dimensional integrable models, like the
existence of a infinite number of conserved charges, can indeed be 
systematically 
generalized to theories in higher 
dimensions.  Our approach is to find the counterpart of the zero 
curvature condition \rf{2dint} in $1+1$ dimensional models. 
It is true that the equation \rf{2dint} is the compatibility or integrability 
condition for  covariant derivatives. 
However, eq.\rf{2dint} is also a conservation law in the sense that the path
ordered exponential of the  connection $A_{\mu}$ along a closed curve is a
constant independent of the curve. That leads to a Gauss type law stating that 
there is a conserved flux through the  curve. We  generalize the 
zero curvature \rf{2dint} to a $d+1$ dimensional space-time by looking for a 
conserved flux through a $d$ dimensional surface, associated to a rank 
$d$ antisymmetric tensor.  Consequently, we are generalizing the concept of
integrability to higher dimensions, from the point of view of conservation laws. 
However, the geometrical and algebraic structures of our approach confirms the
deep connection between  conservation laws and integration of the equations of
motion by leading to interesting methods for constructing large classes of 
solutions.  Let us give a more detailed introduction to these ideas.

Consider a 
connection $A_{\mu}$ and a curve $\G$ on the two dimensional space-time. 
We remind the reader of the construction of conserved quantities. 
Introduce the parallel transport operator $W$ via the equation
\begin{equation}
{d W\o{d\s}} + A_{\mu} {d x^{\mu}\o{d\s}} W = 0
\lab{weqintro}
\end{equation}
where $\s$ parametrizes the curve $\G$.  As we explain in 
Section~\ref{sec:2dint},  eq.~\rf{2dint} is the condition for the 
quantity $W$ to be independent of the curve $\G$ as long as its end 
points are kept fixed.  Therefore, $W$ evaluated on a closed 
contractible curve 
should be equal to unity.  Consequently, as explained in Section 
\ref{sec:2dint}, by taking space-time to be ${\bf R} \times S^1$ for 
instance, any power $N$ of the path ordered exponential $\mbox{Tr} 
\left(P\exp (\int_{S^1} A_x (x,t)dx)\right)^N$ is conserved in time.  
That is how the zero curvature condition \rf{2dint} leads to 
conservation laws.

Our approach in higher dimension is to  generalize the Wilson loop $W$ 
and construct 
quantities that when
integrated over hypersurfaces, they are independent of deformations 
of the hypersurfaces if the boundary is fixed.  We are therefore 
looking for Gauss type laws. The idea is quite simple even though its 
implementation involves several technicalities. Naively, 
in a space-time of dimension $d+1$, one would think of 
introducing just a rank $d$ antisymmetric tensor, and define the counterpart of 
$W$ in \rf{weqintro} as the integration of such tensor on a $d$-dimensional 
surface. However we face two major problems. First we want  
{\em local} conditions for surface independence, i.e., local zero curvature 
conditions. Second we want 
such conditions to be rich enough to be equivalent to the equations of motion 
of nonlinear theories. If we take the rank $d$ tensor to live in an abelian 
algebra then the surface independence follows from the abelian Stokes theorem 
and it  just says that its exterior derivative 
should vanish. This is local but not ``nonlinear enough''. If we take that tensor 
to live in a nonabelian algebra we get highly nonlocal conditions for 
surface independence. Our approach is to introduce, in addition to the 
antisymmetric tensors,  gauge potentials which allows for parallel transport 
along curves on the hypersurfaces. This makes the formulas more local. 
For instance in $2+1$ dimensions, as discussed in Section \ref{sec:3dint},  
we introduce a quantity $V$, depending on an
antisymmetric tensor $B_{\mu\nu}$ and a vector $A_{\mu}$ through the
differential equation
\begin{equation}
{d V\o{d\tau}} - V T(B,A,\tau ) = 0
\lab{veqintro}
\end{equation}
where
\begin{equation}
T(B,A, \tau ) \equiv \int_0^{2\pi} d\s\, W^{-1} B_{\mu\nu} W\,
{d x^{\mu}\o{d\s}} {d x^{\nu}\o{d\tau}}\;,
\lab{tdefintro}
\end{equation}
here $W$ is the same as in \rf{weqintro}.  The quantity $V$ is 
obtained by integrating \rf{veqintro} over a two dimensional surface 
$\Sigma$ with boundary $\G$.  We scan $\Sigma$ with loops starting and 
ending at a fixed point $x_0$ on $\G$.  The loops are parametrized by 
$\s$, and the set of loops scanning $\Sigma$ is parametrized by 
$\tau$.  By imposing $V$ to be independent of deformations of $\Sigma$ 
which keep $\G$ fixed, we get nonlocal conditions on $B_{\mu\nu}$ and 
$A_{\mu}$.  One then has to explore  algebraic structures which yield
local equations. We present two cases  where such 
conditions become local\footnote{There are more complicated scenarios 
but we have not studied them.}.  The first one leads to zero curvature 
conditions for topological models like the BF theory, Chern-Simons and 
$2+1$ gravity, as discussed in Section \ref{subsec::local3d}.  The 
second, which leads to new dynamical integrable systems,  is
\begin{equation}
D_{\mu} {\tilde B}^{\mu} =0 \; , \qquad F_{\mu\nu} =0
\lab{wayoutintro}
\end{equation}
where $D_{\mu}\, \cdot \equiv  \pa_{\mu}\, \cdot + 
\lb A_{\mu} \, , \, \, \cdot \rb $, 
$F_{\mu\nu} =  \lb D_{\mu} \, , \, D_{\nu} \rb$,   and 
${\tilde B}^{\mu} \equiv \h \epsilon^{\mu\nu\rho} B_{\nu\rho}$. In addition, 
$A_{\mu}$ takes value on a Lie algebra $\cg$, ${\tilde B}^{\mu}$  on an
abelian algebra $P$, which transforms under a given representation of $\cg$, 
i.e.
$\lb \cg \, , \, \cg \rb \in \cg$, $\lb \cg \, , \, P \rb \in P$ and 
$\lb P \, , \, P \rb = 0$. The flatness of $A_{\mu}$ is required in order to
have $V$ be independent of the way we choose to scan $\Sigma$ with loops. 

The similarities between our approach and the usual two dimensional curvature 
\rf{2dint} becomes even more apparent if we formulate it in loop space 
in Section~\ref{sect:curv} where we show that equations \rf{wayoutintro}
lead to a flat connection on loop space.

In a space-time of dimension $d+1$ with $d > 2$, there appears some
ambiguities in the sense that we can 
introduce (or not) antisymmetric tensors of rank ranging from $2$ to 
$d$, in addition to the freedom of introducing several connections 
$A_{\mu}$. However, the same reasonings 
lead to local integrability conditions similar to \rf{wayoutintro}.   
Those  ideas are explicit realizations of the potentiality
of the general 
theory of integrability as the vanishing of curvature in higher loop 
spaces presented in Section~\ref{sect:curv}. In Section~\ref{sect:Ex}, 
several examples in three and four 
dimensions are discussed in detail as zero curvature conditions of the form 
\rf{wayoutintro}.  Particular attention is given there to  the $CP^1$ model 
in $2+1$, which  demonstrates the potential usefulness 
of our method in studying integrability in higher dimensions. It is shown 
that it has  relativistic invariant submodels with infinite number of  
new conserved
currents which are given in a simple and systematic way.

The fact that we have quantities which are path independent in loop
space, or consequently  invariant under 
deformations of hypersurfaces, allows us to construct conserved 
charges in a manner similar to the usual two dimensional case.  Such 
quantities are equal to unity when evaluated on a closed contractible 
hypersurface.  The naive idea is then to split such closed 
contractible hypersurfaces into three parts where we have two spatial 
hypersurfaces linked by a temporal hypertube.  The traces of powers of 
such certain quantities integrated over the spatial hypersurface is 
shown to be time independent.  This is shown in detail for the three 
dimensional case.

However, in the case of the local zero curvature conditions 
\rf{wayoutintro} the construction of the conserved currents is quite 
direct.  One has from \rf{wayoutintro} that $J_{\mu} \equiv W^{-1} \, 
{\tilde B}_{\mu} \, W$, satisfies $\pa^{\mu} \, J_{\mu} = 0 $.  
Therefore, the number of conserved currents is equal to the dimension 
of the representation of $\cg$ defined by the abelian algebra where 
${\tilde B}^{\mu}$ lives.  In the case of two dimensional integrable 
models, the infinite number of conservation laws is in general associated to 
infinite dimensional  algebras.  Here, we have a different 
picture.  The infinity of charges comes from infinite dimensional 
representations\footnote{The representation may be reducible.}.  On 
the other hand, it is true that in such cases the nonsemisimple Lie 
algebra formed by $\cg$ and $P$ is infinite dimensional.  In Section 
\ref{subsub:submodel} we discuss a submodel of the $CP^1$ model where 
our method is used to construct an infinite number of nontrivial 
conserved charges.  This result
is easily generalizable to related models like the principal chiral
for any Lie algebra and  nonlinear $\sigma$ models in general. It suggests
in fact a new criterion for integrability in higher dimensions based on 
independence of the representation.

A further  application of local zero curvature conditions of the type
\rf{wayoutintro} is that they open the way
for the development of methods for the construction of solutions. We 
discuss one
possibility which is a type of generalization of the dressing transformation
method in two dimensions. See ref. \ct{savrazu} for an integration scheme using
a different type of zero curvature condition. 
The eqs. \rf{wayoutintro} are invariant under the gauge transformations
\br
A_{\mu}^{(0)} &\ra& A_{\mu} = 
g \, A_{\mu}^{(0)} \, g^{-1} - \pa_{\mu} g \, g^{-1} \nonu\\
B_{\mu\nu}^{(0)} &\ra& B_{\mu\nu} = g \, B_{\mu\nu}^{(0)} \, g^{-1}
+ g\( D_{\mu}^{A^{(0)}} \a_{\nu} - D_{\nu}^{A^{(0)}} \a_{\mu}\) g^{-1}
\lab{dressingintro}
\er
where $g$ is an element of the group obtained by exponentiating the 
algebra $\cg$, and $\a_{\mu}$ are vectors taking values in the abelian 
algebra $P$.  Therefore, if one knows a given solution of the model 
one can obtain new solutions through the transformations 
\rf{dressingintro}.  Just as in the two dimensional dressing method, the 
solution is obtained by equating the potentials written as functionals 
of the fields to the transformed potentials written in terms of the 
parameters of the transformations \rf{dressingintro}.  However, for 
that to work one needs those two sets of potentials to be in the same 
gauge.  In two dimensions there is always a gradation of the algebra 
associated to the problem which guarantees the gauge condition.  Here 
in the higher dimensional cases we do not have necessarily a gradation 
playing that role.  In Section \ref{sec:solcp1}, we apply these ideas 
to construct solutions to the $CP^1$ model.  The problem of the gauge 
fixing of the transformed potentials is made by direct methods and we 
show that the solutions in the orbit of the trivial constant vacuum 
solution are parametrized by the vectors $\a_{\mu}$.  The group 
element $g$ in \rf{dressingintro} is determined by the $\a_{\mu}$'s 
through the gauge fixing conditions. 
The construction of invariant conserved charges, a  criterion of
integrability giving  (sub)models and their infinity of explicit
conserved charges and the generalization of the dressing methods
to obtain solutions, are some of the new results which show the 
potential applications
of the new approach in this paper, which is  organized as follows.

In Section~\ref{sec:2dint} we review the basic aspects of two 
dimensional integrability which we want to generalize to higher 
dimensions.  In Section~\ref{sec:3dint} we discuss the three 
dimensional case with some reference to basic geometric principles in 
loop space.  We derive the conditions for the quantity $V$ to be 
surface independent and explain how to use that to construct conserved 
charges.  As an auxiliary result we obtain useful (local) expressions 
for a generalized Stokes Theorem.  We also show how to obtain local 
zero curvature conditions, and enumerate some examples which are 
easily formulated in our approach, like BF theories, Chern-Simons, 
$2+1$ gravity, the principal chiral model and the $CP^1$ and relevant 
submodels.  In fact the latter is discussed in great detail at the end 
of the paper.  In Section~\ref{sec:4dint}, we discuss the four 
dimensional case.  Here our approach allows the introduction of two 
connections $A_{\mu}$ and ${\cal A}_{\mu}$, and antisymmetric tensors 
$B_{\mu\nu}$ and $H_{\mu\nu\rho}$ in the definition of a quantity $U$ 
integrated over a volume $\Omega$.  We derive the conditions for $U$ 
to be volume independent and discuss the construction of conserved 
charges.  We show that it is possible to obtain basically four types 
of local zero curvature conditions.  In Section~\ref{sect:curv}, we 
discuss how to generalize our approach to any dimension by working in 
higher loop spaces.  We show that our zero curvature condition is 
similar to the two dimensional Zakharov-Shabat equation \rf{2dint} by 
writing it as the flatness condition for a connection defined on the 
space $\Omega^n (M, x_0)$ of maps of an $n$-sphere to a space-time $M$ 
with basepoint $x_0$.  Section~\ref{sect:Ex} is devoted to work out in 
detail some examples: the self-dual Yang-Mills equations and the Bogomolny 
equations in four dimensions, the three dimensional $CP^1$ model in 
great detail, and we also demonstrate the great potential of our method in 
the case of a genuine nontrivial relativistic $2+1$ theory.

\sect{Two dimensional integrability}
\label{sec:2dint}

The concepts and techniques involved in integrable models in one and two 
dimensions are now quite well developed and understood. There exists a variety of
methods for constructing solutions and conserved charges. Certainly a great
deal of the structures related to the concept of integrability are 
particular 
to two dimensions  and cannot perhaps be generalized to higher
dimensions. However, in this section, we want to review some integrability 
concepts which
can indeed be carried on to a space-time of any dimension. The usual zero
curvature  in two dimensions \rf{2dint} 
is the condition for the path ordered integral 
$P\, \exp \( \int_{\Gamma} \, dx^{\mu}\, A_{\mu}\)$, to be independent of the
path $\Gamma$, for fixed end points.  This constitutes some sort of generalized
Gauss law, and it indeed leads to conserved charges as we discuss at the end of
this section. 

As we show in the next sections, such geometrical concepts of 
integrability are the ones that can be implemented in higher 
dimensions.  The conditions for (hyper)surface ordered integrals of 
higher connections to be (hyper)surface independent can be expressed 
as {\it local} equations which constitute generalizations of the zero 
curvature condition \rf{2dint}, in the sense that solutions and local 
conserved charges can be constructed.

The calculations needed to implement those ideas depend crucially on a 
suitable 
generalization of the nonabelian Stokes theorem \ct{stokes} for higher  
order connections, which
to our knowledge has not appeared in the literature.
In this section we rederive the usual version of that theorem for 
ordinary
connections, in a way which will be useful in its generalization to higher
dimensions. 

The calculation that follows is independent of the number of 
dimensions of space-time, and so we will take the greek indices $\mu,
\nu \ldots$ to vary from $0$ to $d$.  Consider a curve $\Gamma$, 
parametrized by $\s$, such that $\s =0$ and $\s = 2 \pi$ correspond to 
the end points of $\G$.  Let $W$ be a quantity defined through the 
differential equation
\begin{equation}
{d W\o{d\s}} + A_{\mu} {d x^{\mu}\o{d\s}} W = 0
\lab{weq} 
\end{equation}
with initial condition $W(0)=I$ 
where $A_{\mu}$ is a connection taking values in the Lie algebra $\cg$ 
of a Lie group $G$. 

Let us now study how $W$ varies under deformations of $\G$ which keep the
initial point $x^{\mu}(\s =0)$ fixed. One gets
\begin{equation}
{d \d W\o{d\s}} + A_{\mu} {d x^{\mu}\o{d\s}} \d W  + 
\d \( A_{\mu} {d x^{\mu}\o{d\s}} \) W= 0
\end{equation}
Multiplying from the left by $W^{-1}$ and using the fact that \rf{weq} implies 
\begin{equation}
{d W^{-1}\o{d\s}} - W^{-1} A_{\mu} {d x^{\mu}\o{d\s}}  = 0 \, , 
\end{equation}
one gets
\begin{equation}
{d \, \o{d\s}}\( W^{-1}\d W \)  = -  
W^{-1} \(  \pa_{\l}A_{\mu} \d x^{\l} {d x^{\mu}\o{d\s}} + 
A_{\mu} {d \d x^{\mu}\o{d\s}}\) W 
\lab{dsw1w}
\end{equation}
After an integration by parts one obtains
\begin{equation}
W^{-1}\d W = - W^{-1}A_{\mu}W \d x^{\mu} + 
\int_0^{\s} d \s^{\pr}W^{-1} F_{\mu\nu} W {d x^{\mu}\o{d\s^{\pr}}}\d x^{\nu}
\lab{varw}
\end{equation}
with
\begin{equation}
F_{\mu\nu} \equiv \pa_{\mu} A_{\nu} - \pa_{\nu} A_{\mu} +
\sbr{A_{\mu}}{A_{\nu}}
\lab{fmunu}
\end{equation}

If we consider variations of $\G$ where the end point at $\s =2\pi$ is also 
kept fixed, then from \rf{varw}
\begin{equation}
W^{-1}(2\pi )\d W(2\pi ) = 
\int_0^{2\pi} d \s^{\pr}W^{-1} F_{\mu\nu} W {d x^{\mu}\o{d\s^{\pr}}}\d x^{\nu}
\lab{var2pi}
\end{equation}

Consider now the case where $\G$ is a closed curve 
($x_0\equiv x^{\mu} (0) = x^{\mu}(2\pi )$), and let $\S$ be a two dimensional 
surface having $\G$ as its boundary.
One can scan $\S$ with loops starting and ending at the fixed point $x_0$, and
such loops can be parametrized by $\tau$ such that $\tau =0$ correspond to the
infinitesimal loop around $x_0$ and $\tau =2\pi$ correspond to $\G$. Then we
can take the variation of $W$ above, to correspond to the deformation of one
loop into the other, i.e. $\d = \d \tau \, {d \, \o d\tau}$. So, we can write 
\rf{var2pi} as 
\begin{equation}
{d W(2\pi )\o{d\tau}} - 
W(2\pi )\, \int_0^{2\pi} d \s^{\pr}W^{-1} F_{\mu\nu} W 
{d x^{\mu}\o{d\s^{\pr}}}{d x^{\nu}\o{d\tau}} = 0
\lab{weq2} 
\end{equation}

The fact that $W$ for a closed loop can be determined either by \rf{weq} or
\rf{weq2} is the very statement of the nonabelian Stokes theorem. Indeed,
integrating those equations one gets
\begin{equation}
P\exp\( \int_{\G} d\s A_{\mu} {dx^{\mu}\o d\s} \) = 
{\cal P} \exp \( \int_{\S} d\tau\,  d\s W^{-1} F_{\mu\nu} W 
{d x^{\mu}\o{d\s}}{d x^{\nu}\o{d\tau}}\)
\lab{stokestheor}
\end{equation}
where $P$ and ${\cal P}$ mean path and surface ordering respectively. Notice
that we have dropped from both sides of \rf{stokestheor} the multiplicative
integration constant which is the initial value of $W$, i.e. $W(x_0)$. 

Conserved quantities are constructed as follows.  First we consider 
the case where spacetime is a cylinder.  At a fixed time $t_0$ 
consider a loop $\gamma_0$ beginning and ending at $x_0$.  At a later 
fixed time $t_1$ consider a loop $\gamma_1$ also beginning and ending 
at $x_0$.  Let $\gamma_{01}$ be a path connecting $(t_0,x_0)$ with 
$(t_1,x_0)$.  The flat connection allows us to integrate the parallel 
transport equation \rf{weq} along two different paths obtaining 
$W(\gamma_0) = W(\gamma_{01})^{-1} W(\gamma_1) W(\gamma_{01})$.  We 
first observe that $W(\gamma_0)$ transforms under a gauge 
transformation $g(x)$ as $W(\gamma_0) \to g(x_0) W(\gamma_0) 
g(x_0)^{-1}$.  The conserved quantity should be gauge invariant.  If $\chi$ 
is a character for the group $G$ we have that $\chi(W(\gamma_0))$ will be 
gauge invariant.  Also $\chi(W(\gamma_0))= \chi(W(\gamma_1))$.  Thus 
we can construct a conserved gauge invariant quantity
\begin{equation}
	\chi(W(\gamma_0)) 
\lab{conserv2d}
\end{equation}
for every independent character of the group. These are the constants 
of motion in the zero curvature construction. Note that the data needed to 
compute $\chi(W(\gamma_0))$ is all determined at time $t_0$.

In the case where the spacetime is two dimensional Minkowski space one 
has to impose physically sensible boundary conditions at spatial 
infinity.  Note that $\;$ $P \exp \left(\int_{-\infty}^\infty A_x dx\right)$ 
is not gauge invariant if one allows nontrivial gauge transformations 
at infinity.  In setting up the problem one has to choose the correct 
physical boundary conditions which may for example require that $A_0$ 
vanishes at infinity.  Depending on the details one can construct a 
conserved quantity by a slight modification of the construction above.

The conserved quantities \rf{conserv2d} are nonlocal because in general the 
connection $A_{\mu}$ lies in a nonabelian algebra. However, in cases like the 
affine Toda models \ct{ot}, it 
is possible to get local conservation laws by gauge transforming $A_{\mu}$ into 
an abelian subalgebra.

\sect{Three dimensional integrability}
\label{sec:3dint}

In analogy with the two dimensional case we will show that the
conservation laws of ``integrable models'' in three dimensions are
associated to surface integrals of a rank two tensor.  Some care must
be taken in order to obtain {\it local} zero curvature conditions.

Consider a two dimensional surface $\S$ with boundary $\G$.  We choose
a basepoint $x_0$ on $\G$, and we scan the surface $\S$ with loops
passing through $x_0$ and being parametrized by $\tau$, such that
$\tau =0$ corresponds to the infinitesimal loop around $x_0$, and $\tau
= 2 \pi$ to the boundary $\G$.  Each loop is
parametrized by $\s$ with $\s =0$ and $\s =2\pi$ corresponding to the
fixed point $x_0$.  We  introduce a rank two antisymmetric tensor
$B_{\mu\nu}$ transforming under the adjoint representation of the group,
a vector potential $A_{\mu}$, and define a quantity $V$ through
the equation
\begin{equation}
{d V\o{d\tau}} - V T(B,A,\tau ) = 0
\lab{veq}
\end{equation}
where
\begin{equation}
T(B,A, \tau ) \equiv \int_0^{2\pi} d\s\, W^{-1} B_{\mu\nu} W\,
{d x^{\mu}\o{d\s}} {d x^{\nu}\o{d\tau}}
\lab{tdef}
\end{equation}
We choose initial condition $V(\tau=0)=I$.
The quantity $W$ depends on the vector $A_{\mu}$ and it is defined through the
equation \rf{weq}.  Geometrically we are parallel transporting $B$ to
$x_0$ so that we can add all the values together. We do not want the
quantity $V$ to
depend upon the way we choose to scan $\S$ with loops passing through $x_0$,
therefore we will impose that the connection $A_{\mu}$ should be flat
\begin{equation}
F_{\mu\nu} \equiv \pa_{\mu} A_{\nu} - \pa_{\nu} A_{\mu} +
\sbr{A_{\mu}}{A_{\nu}} = 0 \;.
\lab{flatamu}
\end{equation}
As we saw in section \ref{sec:2dint}, flatness implies that $W$ is path
independent as long as the end points of the path are kept fixed.
Each point on $\S$ belongs to a given loop of the set scanning
$\S$. Therefore, $W$ is defined on a given point of $\S$ by integrating
\rf{weq} from $x_0$ to that point through the loop it belongs to. If we
change the way we scan $\S$, the value of $W$ associated to  each point of
$\S$ will not change, because the flatness condition \rf{flatamu} guarantees
that
$W$ is path independent because $x_0$ is  kept fixed. Consequently,
the integrand in \rf{tdef} is a local function on $\S$. Therefore, the $V$
resulting from the integration of \rf{veq} is independent of the way we scan
$\S$, because changing the scanning is now equivalent to a change of the
coordinates $( \s , \tau )$ on $\S$ since, as we have shown,
$ W^{-1} B_{\mu\nu} W$ is a function of the points of $\S$ and not of
the loops.

In addition the initial condition $W(x_0)=I$ allows
us to
compute $W$ uniquely on any other point by integrating \rf{weq}. In fact, one
has from \rf{weq}
\begin{equation}
A_{\mu} = - \pa_{\mu} W \, W^{-1}
\lab{flatamuw}
\end{equation}

We now want to study how $V$ changes when we deform $\S$ but keeping the
boundary $\G$ fixed. We will be making variations on the loops scanning
$\S$ which are perpendicular to $\S$. Although we are interested in three
dimensional space-time, the calculations that follow are true in any
dimension.

It follows from \rf{veq} that
\begin{equation}
{d V^{-1}\o{d\tau}} + T(B,A,\tau )\, V^{-1} = 0
\end{equation}
Then varying \rf{veq} one gets
\begin{equation}
{d \, \o{d\tau}}\( \d V \, V^{-1} \)  = V \, \d T(B,A, \tau ) \, V^{-1}
\end{equation}

In calculating the variation of $T(B,A, \tau )$ we make use of \rf{varw} and
\rf{flatamu} to obtain
\begin{equation}
\d W = - A_{\mu}W \d x^{\mu}
\end{equation}
In addition, we perform an integration by parts whenever we have a derivative
of $\d x^{\mu}$ and use the fact that
\begin{equation}
\d x^{\mu}(\s =0 ) = \d x^{\mu}(\s =2\pi ) =
\d x^{\mu}(\tau =0 ) = \d x^{\mu}(\tau =2\pi ) = 0
\lab{endpoints}
\end{equation}
We also use the fact that $W$ satisfies equation \rf{weq} with
$\s$-derivatives replaced by $\tau$-derivatives, since they both lead to
deformations of the contours.

Performing the calculations one obtains
\br
\d V(\tau ) \, V^{-1}(\tau )&=&
V(\tau )\( \int_0^{2\pi} d\s \, \, W^{-1} B_{\mu\nu} \, W\,
{d x^{\mu}\o{d\s}} \, \d x^{\nu}\) \, V^{-1}(\tau )\nonu\\
&+& \int_0^{\tau} d\tau^{\pr} \, V(\tau^{\pr})\nonu\\
&\times& \( \int_0^{2\pi} d\s \,
W^{-1} \( D_{\l} B_{\mu\nu} + D_{\mu} B_{\nu\l} +  D_{\nu} B_{\l\mu} \) W  \,
{d x^{\mu}\o{d\s}} {d x^{\nu}\o{d\tau^{\pr}}} \d x^{\l}\right.  \nonu\\
 &-&\left.  \lb T(B,A, \tau^{\pr} ) \, , \,
\int_0^{2\pi} d\s \, W^{-1} B_{\mu\nu} \, W \, {d x^{\mu}\o{d\s}} \d x^{\nu}
\rb \) \, V^{-1}(\tau^{\pr})
\lab{varv0}
\er
where
\begin{equation}
D_{\l} B_{\mu\nu} \equiv \pa_{\l} B_{\mu\nu} + \lb A_{\l} \, , \, B_{\mu\nu}\rb
\lab{covder}
\end{equation}
Notice that if we set $\tau=2\pi$   in \rf{varv0}
one gets from \rf{endpoints} that
the first term on the r.h.s. of \rf{varv0} vanishes.

Consider now the case where $\S$ is a closed surface where the loop
$\Gamma$ has collapsed to
the fixed point $x_0$. Let $\O$ be a volume whose boundary is
$\S$. Analogously, we can scan the volume $\O$ with closed surfaces which have
the point $x_0$ in common. Such surfaces can be parametrized by $\zeta$ such
that $\zeta =0$ corresponds to the infinitesimal surface around $x_0$ and
$\zeta = 2\pi$ to the boundary $\S$. Then we can consider the variation of $V$
that corresponds to the deformation of one closed surface to the other.
We can write ($\d \equiv d\zeta {d \, \o d\zeta}$)
\begin{equation}
{d V_c\o{d\zeta}} - \( \int_0^{2\pi} d\tau \, V\, {\cal K}\,
V^{-1}\)\, V_c  = 0
\lab{veq2}
\end{equation}
where $V_c$ stands for $V$ defined on a closed surface, and
\begin{equation}
{\cal K}\equiv
 \int_0^{2\pi} d\s \, W^{-1} \(
D_{\l} B_{\mu\nu} + D_{\mu} B_{\nu\l} +  D_{\nu} B_{\l\mu} \) W  \,
{d x^{\mu}\o{d\s}} {d x^{\nu}\o{d\tau}} {d x^{\l}\o {d\zeta}}
 -  \lb T(B,A, \tau ) \, , \,  T(B,A, \zeta) \rb
\lab{fdef}
\end{equation}

The fact that $V$ for closed surfaces can be obtained by integrating either
\rf{veq} or \rf{veq2} leads us to the formulation of a generalized nonabelian
Stokes theorem. The integration of \rf{veq} and \rf{veq2} gives
\begin{equation}
{\cal P} \exp \(  \int_{\S}d\tau d\s\, W^{-1} B_{\mu\nu} W\,
{d x^{\mu}\o{d\s}} {d x^{\nu}\o{d\tau}}\) =
{\hat {\cal P}} \exp \( \int_{\O} d\zeta d\tau \,  V\, {\cal K}\, V^{-1}\)
\end{equation}
where ${\cal P}$ and ${\hat {\cal P}}$ mean surface and volume ordering
respectively.

This expression has many applications.  We will use it here to 
construct conserved charges and to discern integrable local field 
theories.  As we will show in Section~\ref{sect:curv}, $\cal K$ is in 
fact a curvature in the space of loops corresponding to the connection 
$T$ in eq~\rf{tdef}.

\subsection{Construction of conserved charges}
\label{sect:charges-3d}

Constructing the conserved quantities in $(2+1)$-dimensions is more
complicated and more subtle. Naive idea is basically correct except
that one has to be very careful and verify that everything is gauge
invariant and only depends on data at the fixed time slice. For
simplicity we take spacetime to be ${\bf R}\times S^2$ to avoid
intricate model dependency on boundary conditions at spatial infinity.

\begin{figure}[tbp]
	\centerline{\epsfxsize=0.25\textwidth\epsfbox{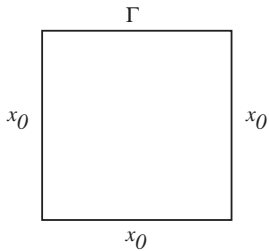}}
	\caption[squares]{\parbox[t]{.7\textwidth}{The left, bottom and right
	edges of the square get mapped to $x_0$ and the top edge is mapped
	to $\Gamma$.}}
	\protect\label{fig:square}
\end{figure}

Consider a surface $\Sigma$ with boundary $\Gamma$ and basepoint $x_0
\in \Gamma$.  We can think of such a surface as a map $f$ from the
square $[0,2\pi]^2$ parametrized by $(\sigma,\tau)$ to the spacetime
with the following constraints.  The left, bottom and right edges get
mapped to $x_0$ and the top edge maps to the curve $\Gamma$ which is
the point set $\{f(\sigma,2\pi) | \sigma\in [0,2\pi]\}$.  We will
represent this pictorially as Figure~\ref{fig:square}.
To such a
surface we associate $V(\Sigma,x_0)$ by integrating (\ref{eq:veq}).  One
can verify that under a gauge transformation $g(x)$ we have
$V(\Sigma,x_0) \to g(x_0)^{-1} V(\Sigma,x_0) g(x_0)$.  We are
ultimately interested is allowing at a fixed time $\Sigma$ to wrap
once around $S^2$ while $\Gamma$ collapses to $x_0$.  This
$V(S^2,x_0)$ is not gauge invariant transforming via the adjoint
representation at $x_0$. We can take a character which will be
gauge invariant.  Our conserved quantities are $\chi(V(S^2,x_0))$
though we have not proved this yet.  In fact the invariant is actually
independent of the choice of basepoint.  To see this go back to the
definition of $T$ and ask what happens if you change basepoint to
$x_0'$ elsewhere on the boundary.  The first observation is that
parallel transport from $x_0$ to $x_0'$ is independent of path, i.e.,
it is only a function $W(x_0',x_0)$ of the two points on $\Gamma$.
You can compare the definition of the two $T$'s with the two choices
of basepoints and see that they are related via conjugation by
$W(x_0',x_0)$.  Since the conjugation element is independent of
$(\sigma,\tau)$ one can immediately see from differential equation
(\ref{eq:veq}) how changing the basepoint affects $V$.  The conclusion is
that
$$
V(\Sigma,x_0') = W(x_0',x_0)^{-1}V(\Sigma,x_0)W(x_0',x_0)\;.
$$
Thus characters of $V$ are independent of the choice of basepoint.
Our tentative constant of the motion $\chi(V(S^2,x_0))$ does not
depend on the basepoint.  We now have to prove that it is a constant
of motion.

\begin{figure}[tbp]
	\centerline{\epsfysize=0.25\textheight\epsfbox{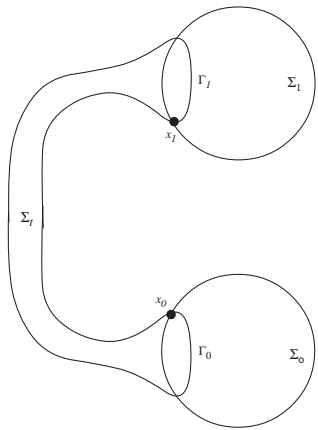}}
	\caption[retort]{\parbox[t]{.7\textwidth}{This figure may be viewed
	as the union of a top disk $\Sigma_1$ with boundary $\Gamma_1$, a
	cylinder $\Sigma_t$ with boundary $\Gamma_1\cup\Gamma_0$ and a
	bottom disk $\Sigma_0$ with boundary $\Gamma_0$ all glued along the
	common boundaries. We denote $\Sigma_t\cup\Sigma_1$ by
	$\widetilde{\Sigma}_0$.}}
	\protect\label{fig:retort}
\end{figure}
Consider the situation depicted in Figure~\ref{fig:retort} of a closed
surface $\Sigma_0 \cup \widetilde{\Sigma}_0$ made by joining along a
common boundary $\Gamma_0$.  We assign the same basepoint $x_0$ to
both surfaces.  The surface $\Sigma_0 \cup \widetilde{\Sigma}_0$ is
homotopically trivial in ${\bf R}\times S^2$.  By choosing
orientations correctly on the surfaces it is clear that
\begin{equation}
	V(\Sigma_0,x_0) = V(\widetilde{\Sigma}_0, x_0)
	\label{Vsigma}
\end{equation}
because we proved that with $F=0$ and $DB=0$ the value of $V$ is
independent of $\Sigma$ as long as $\partial\Sigma=\Gamma_0$.  To get
a constant of the motion we would like to proceed as follows.  Write
$\widetilde{\Sigma}_0 = \Sigma_t \cup \Sigma_1$.  Assume $\Sigma_0$
and $\Sigma_1$ are fixed time surfaces at respective times $x_0^0$ and
$x_1^0$.  Collapse the respective boundaries $\Gamma_0$ and $\Gamma_1$
to points $x_0$ and $x_1$ so that the respective closed surfaces wrap
once around the ``spatial sphere''.  Now collapse $\Sigma_t$ to a
curve.  Since $V$ is roughly a surface integral we would naively
expect the contribution from $\Sigma_t$ to vanish and conclude
$V(S^2,x_0) = V(S^2,x_1)$.  This naive answer  does not have the
right gauge transformation properties.  In fact, the correct answer as
we will soon show is that
\begin{equation}
	V(S^2,x_0) = W(x_1,x_0)^{-1} V(S^2,x_1)W(x_1,x_0)
	\label{V-eq}
\end{equation}
where $W(x_1,x_0)$ is parallel transport from $x_0$ to $x_1$.  This
equation transforms correctly under gauge transformations.  By taking
a character we have $\chi(V(S^2,x_0))=\chi(V(S^2,x_1))$.  Therefore
\begin{equation}
	\chi(V(S^2,x_0))
	\label{eq:char}
\end{equation}
is a constant of the motion.  By previous arguments both sides are
independent of the basepoints.  Thus we have constructed constants of
motion which only depend on data at fixed time.  The constant of
motion does not depend on choice of basepoint.

We now begin proving (\ref{V-eq}).  The easiest way to solve for $V$
is to exploit the ability to integrate differential equation
(\ref{eq:veq}) backwards.  We will begin with the loop $\Gamma_0$ and
integrate to the point $x_0$.  Our surface of interest is
$\widetilde{\Sigma}_0$ which is topologically a disk.
\begin{figure}[tbp]
	\centerline{\epsfysize=0.25\textheight\epsfbox{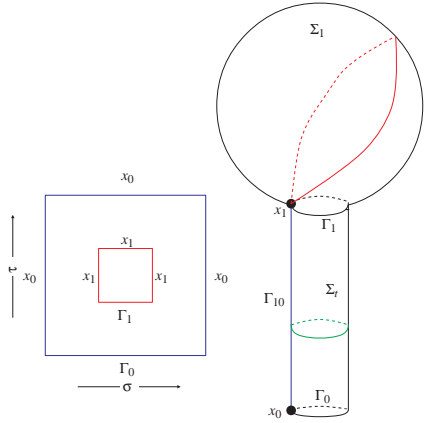}}
	\caption[cap]{\parbox[t]{.7\textwidth}{%
	On the left we have the representation of the scan of
	$\widetilde{\Sigma}_0$ in the $(\sigma,\tau)$ square. On the right,
	all loops
	begin at $x_0$. We draw a loop with $0<\tau< 1/3$ and another
	one with $1/3 < \tau< 2/3$.
	}}
	\protect\label{fig:cap}
\end{figure}
For our purposes it is best to think of a disk as a cylinder with an
end cap.  We scan the surface as in Figure~\ref{fig:cap}.  On the
left we have the representation
in the $(\sigma,\tau)$ square.
To make the discussion clearer we will
rescale the parameters $\sigma$ and $\tau$ such that they range from
$0$ to $1$.  Fix a path $\Gamma_{10}$ from $x_0$ to $x_1$.  By making
good choices for paths of integration we can simplify the computation.

For $0 \le \tau \le 1/3$, the integration curve $\gamma_\tau$ needed to
compute $T$ is the following.  Move upwards along $\Gamma_{10}$ until at
parameter $\sigma=1/3$ you are fraction $3\tau$ to $x_1$.  Now move
circumferentially until you return to the starting point of the
circle.  This should now be $\sigma = 2/3$.  Finally move downwards
along $\Gamma_{10}$ until you return to $x_0$ at $\sigma=1$.  We increase
$\tau$ until we get to $\tau=1/3$ where the circumferential path is
$\Gamma_1$.

For $\tau\in (1/3,2/3]$ we choose the curve $\gamma_\tau$ in the
following way. Begin at $x_0$ and move along $\Gamma_{10}$ such that at
$\sigma=1/3$ you are at $x_1$. For $\sigma \in [1/3,2/3]$ you are in a
path which is no different than one which you will use to scan the
cap $\Sigma_1$. You require that the path ends at $x_1$ at $\sigma=2/3$.
Finally you move down along $\Gamma_{10}$ until at $\sigma=1$ you are at $x_0$.
At $\tau=2/3$ you have finished scanning the cap and the path
$\gamma_{2/3}$ is just $\Gamma_{10}$ traversed forwards and backwards.

Finally for $\tau \in (2/3,1]$ what you do is begin at $x_0$ move up
along $\Gamma_{10}$ a fraction $3(1-\tau)$ to $x_1$ and reverse your path.
This is the interpretation of Figure~\ref{fig:cap}.

We have to integrate differential equation (\ref{eq:veq}). We first
integrate from $\tau=0$ to $\tau=1/3$ obtaining $V(\Sigma_\tau)$. We
do not need to know much about $V(\Sigma_\tau)$. Continue integrating
the differential equation. Let us study the form of (\ref{eq:tdef}) for
$\tau\in [1/3,2/3]$. The contribution to integral (\ref{eq:tdef})
for $\sigma\in [0,1/3]$ cancels the contribution for $\sigma\in
[2/3,1]$ because the path $\Gamma_{10}$ has been traversed forward
and backwards. We are left with
$$
T_{x_0}(\tau ) = \int_{1/3}^{2/3} d\sigma\, W(x,x_0)^{-1}
B_{\mu\nu}(x) W(x,x_0)\,
{d x^{\mu}\over{d\sigma}} {d x^{\nu}\over{d\tau}}
$$
where we have used the notation that $W(z,y)$ is parallel transport
from $y$ to $z$.  The notation $T_{x_0}$ refers to that we are
parallel transporting back to $x_0$.  In the above $x$ represents
$x(\sigma,\tau)$.  We simply observe that $W(x,x_0) =
W(x,x_1)W(x_1,x_0)$.  Inserting this into the above leads to
$$
T_{x_0}(\tau ) = W(x_1,x_0)^{-1}
\left(\int_{1/3}^{2/3} d\sigma\, W(x,x_1)^{-1}
B_{\mu\nu}(x) W(x,x_1)\,
{d x^{\mu}\over{d\sigma}} {d x^{\nu}\over{d\tau}}\right)
W(x_1,x_0)
$$
Notice that what is between the large parentheses is precisely
$T_{x_1}$ which is what we need to calculate $V(\Sigma_1,x_1)$. As
$\tau$ grows to $2/3$ we scan the cap $\Sigma_1$. Thus if we
return to the differential equation and integrate to $\tau=2/3$ we
obtain  $V(\Sigma_t) W(x_1,x_0)^{-1} V(\Sigma_1,x_1)$ $W(x_1,x_0)$.
The integration from $\tau=2/3$ to $\tau=1$ is trivial since $T$
vanishes for that range.
Our final result is that
$$
	V(\widetilde{\Sigma}_0,x_0) =
	V(\Sigma_t) W(x_1,x_0)^{-1}V(\Sigma_1,x_1)W(x_1,x_0)
$$
Using deformation independence we have
\begin{equation}
	V(\Sigma_0,x_0) =
	V(\Sigma_t) W(x_1,x_0)^{-1}V(\Sigma_1,x_1)W(x_1,x_0)
	\label{V-true}
\end{equation}
This is our main result which may also
be used to prove the Lorentz invariance of our charges.  We  use
a special case to establish (\ref{V-eq}) by observing that collapsing
$\Sigma_t$ to a curve results in $V(\Sigma_t)=I$ because $T=0$ along
the collapsed curve.

\subsection{Local integrability conditions}
\label{subsec::local3d}

As we have seen, in order to construct conserved quantities we have to 
require the quantity $V$, introduced in \rf{veq}, to be surface 
independent.  A sufficient condition is to have the curvature 
${\cal K}$, defined in \rf{fdef}, equal to zero.  Note that ${\cal K}$ 
is a nonlocal quantity.

We want to discuss now two possible ways of introducing local 
conditions that guarantee the vanishing of ${\cal K}$.  These will 
constitute local generalizations, in three dimensions, of the zero 
curvature condition \rf{2dint} in two dimensions.  We will construct 
examples of higher dimensional field theories where these conditions 
are generically equivalent to or  imply the classical equations of 
motion.  This distinction will be at the basis of an integrability 
criterion, which can in turn unravel new integrable systems.  
Furthermore, 
it will be shown how our geometrical formulation can be used to 
construct solutions and conserved charges, an infinite number in the 
integrable case.

\subsubsection{First type of integrable models: BF Theories, Chern-Simons and
$2+1$ gravity}

We have taken the connection $A_{\mu}$ to be flat in order
for $V$ to be independent of the way we scan the surface with loops. We now
consider the
cases where $B_{\mu\nu}$ is covariantly constant
\begin{equation}
D_{\l} B_{\mu\nu} = 0 \; , \qquad F_{\mu\nu} =0
\lab{bfeq}
\end{equation}
Now, if we choose the value of $B_{\mu\nu}$ on a given point of space time
to be, let us say, $B_{\mu\nu}^{(0)}$ then
\begin{equation}
B_{\mu\nu} (x) \equiv W(x) B_{\mu\nu}^{(0)} W^{-1}(x)
\lab{covb}
\end{equation}
satisfies \rf{bfeq} by virtue of \rf{flatamuw}. In addition, from \rf{tdef}
\begin{equation}
T(B,A, \tau ) = B_{\mu\nu}^{(0)} \, \int_0^{2\pi} d\s\,
{d x^{\mu}\o{d\s}} {d x^{\nu}\o{d\tau}}
\end{equation}
So, if all the components of $B_{\mu\nu}^{(0)}$
commute among themselves, i.e.
\begin{equation}
\lb B_{\mu\nu}^{(0)} \, , \,  B_{\rho\s}^{(0)} \rb =0
\lab{commb}
\end{equation}
one concludes from \rf{fdef} and \rf{bfeq} that ${\cal K}$ vanishes. Notice
that \rf{covb} and \rf{commb} imply that on every point of space time, all
components of $B_{\mu\nu}$ commute among themselves, but components at
different points do not have to do so.

Therefore, from the arguments leading to (\ref{eq:char}) one concludes
that we
have conserved quantities. In three space time dimensions, introduce the
constant Lie algebra valued vector
\begin{equation}
v^{\mu} \equiv \h \, \varepsilon^{\mu\nu\rho}\, B_{\nu\rho}^{(0)}
\end{equation}
Then
\begin{equation}
\int_{\S}d\tau d\s\, W^{-1} B_{\mu\nu} W\,
{d x^{\mu}\o{d\s}} {d x^{\nu}\o{d\tau}} =
v^{\rho}\, \varepsilon_{\rho\mu\nu} \,
\int_{\S}d\tau d\s\,{d x^{\mu}\o{d\s}} {d x^{\nu}\o{d\tau}} \equiv
v^{\rho}\, S_{\rho}^{\S}
\end{equation}
where $S_{\rho}^{\S}$ is the vector associated to the area of the
surface $\S$.
We get the conservation laws (\ref{eq:char}) given by $\chi\left(\exp
v^\mu S_\mu(S^2)\right)$.

The equations of motion of the BF theory \ct{bf},  without the kinetic terms,
are just \rf{bfeq}. Therefore the sector of these theories, corresponding to
those
$B_{\mu\nu}$ constructed out of an abelian $B_{\mu\nu}^{(0)}$, constitute an
example of an ``integrable'' theory in $2+1$ dimensions. Of course, the
classical conservation law is trivial because we are dealing with a
``topological'' theory.
The equations of motion of the Chern-Simons theory are just $F_{\mu\nu} =0$,
and so these models are also examples of our construction. We can also include
$2+1$ gravity into the game, since it is equivalent to the Chern-Simons theory
for the $2+1$ Poincar\'e algebra \ct{2p1}.

\subsubsection{The second type of integrable models}
\label{sec:secondtype}

Let ${\hat{\cal G}}$ be a nonsemisimple Lie algebra and ${\cal P}$ an abelian
ideal of ${\hat{\cal G}}$. Then, if one takes 
\begin{equation}
A_{\mu} \in {\hat{\cal G}} \; ; \qquad B_{\mu\nu} \in {\cal P}
\end{equation}
it follows that
\begin{equation}
W^{-1}B_{\mu\nu}W \in {\cal P}
\lab{incalp}
\end{equation}
and so, the commutator term in \rf{fdef} vanishes
\begin{equation}
\lb T(B,A, \tau ) \, , \,  T(B,A, \zeta) \rb = 0
\lab{ttcomm}
\end{equation}
since ${\cal P}$ is abelian.

Therefore, if we now impose
\begin{equation}
D_{\l} B_{\mu\nu} + D_{\mu} B_{\nu\l} +  D_{\nu} B_{\l\mu} = 0
\lab{covconserb}
\end{equation}
we get, from
\rf{fdef}, that ${\cal K}=0$ and so from \rf{varv0} we have  that
$\d V \, V^{-1} =0$.
Therefore, $V$ is independent of the surface $\S$ as long as its border is
kept fixed.

In three space time dimensions we introduce the dual of $B_{\mu\nu}$ as
\begin{equation}
{\tilde B}^{\mu} \equiv \h \epsilon^{\mu\nu\rho} B_{\nu\rho}
\lab{bdual}
\end{equation}
Therefore, the conditions for $V$ to be surface independent are written as
(\rf{covconserb} and the flatness of $A_{\mu}$ \rf{flatamu})
\begin{equation}
D_{\mu} {\tilde B}^{\mu} =0 \; , \qquad F_{\mu\nu} =0
\lab{wayout}
\end{equation}
These are the local integrability conditions which we introduce for theories
defined on three dimensional space time,
and which constitute a generalization of the zero curvature condition
\rf{2dint} in two dimensions. Notice therefore, that the appearance of 
local integrability
conditions is associated to the structures of nonsemisimple Lie algebras. 
Principal 	chiral models for any Lie algebra, nonlinear sigma models  and
 $CP^1$ and submodels are
easily formulated in this scheme. We will  discuss the latter
in detail at the end.

According to Levi's theorem \ct{jacobson} it follows that if 
${\hat{\cal G}}$ is finite-dimensional and ${\cal P}$ is its radical 
(maximal solvable ideal), then ${\hat{\cal G}} = {\cal H} + {\cal P}$, 
where ${\cal H}$ is a semi-simple Lie subalgebra of ${\hat{\cal G}}$.  
The examples we shall treat in this paper are of this type, and 
therefore we will consider nonsemisimple Lie algebras of the following 
form.  Let $\cg$ be a Lie algebra and $R$ be a representation of it.  
We then construct the nonsemisimple Lie algebra $\cg_R$ as
\br
\lb T_a \, , \, T_b \rb &=& f_{ab}^c T_c \nonu\\
\lb T_a \, , \, P_i \rb &=& P_j R_{ji}\( T_a\) \nonu\\
\lb P_i \, , \, P_j \rb &=& 0
\lab{rt}
\er
where $T_a$ constitute a basis of $\cg$ and $P_i$ a basis for the abelian
ideal $P$ (representation space). The fact that $R$ is a matrix
representation, i.e.
\begin{equation}
\lb R\( T_a \) \, , \, R\( T_b\)  \rb =  R\( \lb T_a \, , \, T_b \rb \)
\lab{rep}
\end{equation}
follows from the Jacobi identities.

We take the connection $A_{\mu}$ to be in $\cg$ and the antisymmetric tensor
field $B_{\mu\nu}$ to be in $P$, i.e.
\begin{equation}
A_{\mu} = A_{\mu}^a T_a \; , \qquad B_{\mu\nu} = B_{\mu\nu}^i P_i
\end{equation}
Therefore, all the results obtained above, i.e. \rf{incalp}-\rf{wayout}, apply
equally well to this particular case. 

It is worthwhile to think a bit about the structure of the Lie
algebra $\cg_R$. The $\cg$-representation $R$ acts on a vector space
$V_R$. Mathematically we have that $V_R$ is an abelian ideal in
$\cg_R = \cg \oplus V_R$. It may be that there is more than one choice
of representation $R$ which makes \rf{wayout} valid. For example
assume we have representations $\{R_1,R_2,\ldots,R_n\}$ and
representation spaces $\{V_1,\ldots,V_n\}$ in which our
formulation is valid. One can now consider this as either a problem
involving $n$ Lie algebras $\{\cg_{R_1},\ldots,\cg_{R_n}\}$ or a
problem involving a single big Lie algebra $\cg_{\rm big} = \cg \oplus
V_1 \ldots \oplus V_n$ where the abelian ideals $V_j$ commute among
themselves. In Section~\ref{subsub:submodel} we construct an infinite
number of conserved currents for a certain model based on finding an
infinite sequence of representations $\{R_1,R_2,R_3,\ldots\}$. As
just mentioned one may think of the construction of this infinity of
conserved currents as due to the presence of an infinite dimensional
Lie algebra. In fact, as discussed in the next subsection, this can
be taken as a criterion for integrability.

Notice that equations \rf{wayout} are indeed conservation laws.
It follows from \rf{flatamuw} and \rf{wayout} that the currents
\begin{equation}
J_{\mu} \equiv W^{-1} \, {\tilde B}_{\mu} \, W
\lab{invcur}
\end{equation}
are conserved
\begin{equation}
\pa^{\mu} \, J_{\mu}  = 0
\lab{invcurcons}
\end{equation}

The conserved charges obtained from \rf{invcurcons} are the same as those from
(\ref{eq:char}). Indeed, since $B_{\mu\nu}$ lives on an abelian invariant
subalgebra, the surface ordering is not necessary:
\begin{equation}
\int_{S^2}  d\tau d\s \, W^{-1} B_{\mu\nu} W\,
{d x^{\mu}\o{d\s}} {d x^{\nu}\o{d\tau}} =
\int_{S^2} ds_{\mu} \, J^{\mu}
\end{equation}
where $ds_{\mu} \equiv
d\tau d\s \epsilon_{\mu\nu\rho}{d x^{\nu}\o{d\s}} {d x^{\rho}\o{d\tau}}$.

Notice that \rf{wayout} is invariant under the local gauge transformations
\br
A_{\mu} &\ra & A^{g}_{\mu} =
g \, A_{\mu} \, g^{-1} - \pa_{\mu} g \, g^{-1} \nonu\\
B_{\mu\nu} &\ra & B_{\mu\nu}^{g} = g \, B_{\mu\nu} \, g^{-1} \,
\qquad \qquad  \qquad \qquad
\mbox{\rm with $g \equiv \exp  \cg $}
\lab{gauge}
\er

The currents \rf{invcur} are gauge invariant.  The transformations
\rf{gauge} imply that $W(x) \ra g(x) \, W(x) g(x_0))^{-1}$ where we
have chosen to take $x_0 = \infty$.  We require $g(\infty)=1$ and
consequently we find that $J_{\mu} \ra J_{\mu}$.

Since the connection $A_{\mu}$ is flat, i.e.
$\lb D_{\mu} \, , \, D_{\nu}\rb = 0$, we have that \rf{wayout} are also
invariant under the gauge  transformations
\br
A_{\mu} &\ra & A_{\mu} \nonu\\
B_{\mu\nu} &\ra & B_{\mu\nu} + D_{\mu} \a_{\nu} - D_{\nu} \a_{\mu}
\lab{newgauge}
\er
where $\a_{\mu}$ is a vector parametrizing the transformations.

Under \rf{newgauge}, the currents \rf{invcur} transform as
\begin{equation}
J_{\mu} \ra J_{\mu} + \epsilon_{\mu\nu\rho} \pa^{\nu}\( W^{-1} \a^{\rho}W\)
\lab{notinvcur}
\end{equation}

The algebra of the gauge transformations \rf{gauge} is isomorphic to the
algebra of the $T_a$'s. The algebra of \rf{newgauge} is abelian, and in
fact the
same as the algebra of the $P_i$'s. The transformations \rf{gauge} and
\rf{newgauge} do not commute.  Indeed, performing \rf{gauge}
first and then \rf{newgauge} one gets
\br
A_{\mu} &\ra &
g \, A_{\mu} \, g^{-1} - \pa_{\mu} g \, g^{-1} \nonu\\
B_{\mu\nu} &\ra &  g \, B_{\mu\nu} \, g^{-1}
+ D_{\mu}^{A^g} \a_{\nu} - D_{\nu}^{A^g} \a_{\mu}\,
\lab{gauge12}
\er
Now, performing \rf{newgauge} first and then \rf{gauge} one gets
\br
A_{\mu} &\ra &
g \, A_{\mu} \, g^{-1} - \pa_{\mu} g \, g^{-1} \nonu\\
B_{\mu\nu} &\ra &  g \, B_{\mu\nu} \, g^{-1}
+ g\( D_{\mu}^{A} \a_{\nu} - D_{\nu}^{A} \a_{\mu}\) g^{-1}
\lab{gauge21}
\er
And one can check that
\begin{equation}
 D_{\mu}^{A^g} \a_{\nu} = g\( D_{\mu}^{A} \( g^{-1}\a_{\nu}g\) \) g^{-1}
\end{equation}

In fact, taking infinitesimal transformations, with $g \sim 1 + \epsilon$
($\epsilon \equiv \epsilon^a T_a$), one gets that
\br
\lb \d_{\epsilon}\, , \, \d_{\a} \rb A_{\mu} &=& 0
\lab{algebratransf} \\
\lb \d_{\epsilon}\, , \, \d_{\a} \rb B_{\mu\nu} &=&
D_{\mu}^{A} \lb \epsilon \, , \,  \a_{\nu} \rb -
D_{\nu}^{A} \lb \epsilon \, , \,  \a_{\mu} \rb \nonu
\er
Therefore, the full algebra of gauge transformations \rf{gauge} and
\rf{newgauge} is isomorphic to the algebra $\cg_R$ \rf{rt}.

These gauge transformations play an important role in the construction of
solutions. In fact, if one knows a (trivial) solution for the zero curvature
conditions \rf{wayout}, then the transformed potentials $A_{\mu}$ and
$B_{\mu\nu}$ correspond to a new (non trivial) solution. That enable us to
implement a dressing like method for the construction of solutions. We shall
discuss that on a more concrete basis in the case of the $CP^1$ model in
section
\ref{sec:solcp1}.

\vspace{1 cm}

\sect{Four dimensional integrability}
\label{sec:4dint}

Building on the results of the two and three dimensional cases we now 
consider a three dimensional volume $\O$ with boundary $\pa\O$.  We 
choose a fixed point $x_0$ on $\pa\O$ and we scan the volume $\O$ with 
closed two dimensional surface, all having basepoint $x_0$ in common.  
We parametrize these closed surfaces by parameter $\zeta$ such that 
$\zeta =0$ corresponds to the infinitesimal closed surface around 
$x_0$ and $\zeta = 2\pi$ corresponds to the boundary $\pa\O$.  We 
then introduce\footnote{The formulation below is actually 
more general than what is needed in the examples of $4$-dimensional 
integrability. This more general structure reduces to the 
simpler structure discussed in Section~\ref{sect:higher-loop} in our 
examples.}
antisymmetric tensors $H_{\mu\nu\rho}$ and 
$B_{\mu\nu}$, and two vectors $A_{\mu}$ and $\ca_{\mu}$.  We define 
the quantity $U$ through the equation
\begin{equation}
{d U \o {d \zeta}} + S(H,B,A, \ca ,\zeta ) \, U = 0
\lab{ueq1}
\end{equation}
where
\begin{equation}
S(H,B,A, \ca ,\zeta ) =  \int_{0}^{2\pi} d\tau \, V \, \( 
\int_{0}^{2\pi} d\sigma \,
\cw^{-1}\, H_{\mu\nu\rho} \cw \, {dx^{\mu}\o{d\sigma}} 
{dx^{\nu}\o{d\tau}}{dx^{\rho}\o{d\zeta}}\) \, V^{-1}
\lab{rdef1}
\end{equation}
and where $V$ is defined in \rf{veq}, and $\cw$ satisfies the 
equation 
\rf{weq} with $A_{\mu}$ replace by  $\ca_{\mu}$, i.e. 
\begin{equation}
{d \cw\o{d\s}} + \ca_{\mu} {d x^{\mu}\o{d\s}} \cw = 0
\lab{cweq} 
\end{equation}

The quantity $S(H,B,A, \ca ,\zeta )$ is defined on every closed 
surface scanning $\O$, and in order to calculate it we scan every one 
of these closed surfaces with loops passing through the common fixed 
point $x_0$.  We parametrize such scanning by $\tau$, such that $\tau 
=0$ and $\tau = 2\pi$ correspond to infinitesimal loops around $x_0$, 
defining the beginning and end of the scanning.  All these loops are 
parametrized in their turn by $\s$, such that $\s =0$ and $\s = 2\pi$ 
correspond to $x_0$, their initial and final point.

Although the actual integration of the equation \rf{ueq1} involves a 
scanning of the volume $\O$, we do not want $U$ to depend upon the 
choice of that scanning.  Now, in any scanning each point of the 
volume $\O$ belongs to a given loop, which in its turn belongs to a 
given closed surface.  The quantity $\cw^{-1}\, H_{\mu\nu\rho} \cw$ in 
\rf{rdef1} is defined on each point of $\O$, but its actual value is 
found by integrating \rf{cweq}, from $x_0$ to that given point, 
through the loop that such point belongs to.  However, as we discussed 
below \rf{flatamu}, the quantity $\cw^{-1}\, H_{\mu\nu\rho} \cw$ will 
depend only on the point and not on the loop that the point belongs to. 
If the connection $\ca_{\mu}$ is flat then one has
\begin{equation}
\flie_{\mu\nu} \equiv \pa_{\mu} \ca_{\nu} - \pa_{\nu} \ca_{\mu} +
\sbr{\ca_{\mu}}{\ca_{\nu}} = 0
\lab{flatcamu}
\end{equation}

The quantity
\begin{equation}
dX(\tau ) \equiv 
\int_{0}^{2\pi} d\sigma \,
\cw^{-1}\, H_{\mu\nu\rho} \cw \, {dx^{\mu}\o{d\sigma}} 
dx^{\nu}dx^{\rho}
\lab{xtau}
\end{equation}
is a function of the loop labeled by $\tau$, and does not depend 
upon the 
closed surface to which such loop belongs to. However, the quantity 
$V(\tau )$ depends not only on the loop labeled by $\tau$ but also, 
according to \rf{veq}, on the surface it was integrated through. 
Therefore, 
if we want the quantity 
\begin{equation}
V(\tau )\, dX(\tau )\, V^{-1}(\tau )
\end{equation}
to depend only on the loop, we have to impose that $V$ should be 
invariant 
under deformations of the surface which keep the boundary fixed. But, 
from 
\rf{varv0} and \rf{fdef} we saw that a sufficient condition for that is
\begin{equation}
{\cal K}(B,A) = 0
\lab{zerok}
\end{equation}
   
Therefore \rf{flatcamu} and \rf{zerok} are sufficient conditions for 
the 
quantity $U$ defined in \rf{ueq1}, to be independent of the way we 
scan $\O$. 
According to \rf{varw} and \rf{varv0} we then have that
\begin{equation}
\d \cw = - \ca_{\mu}\cw \d x^{\mu}
\lab{varcwflat}
\end{equation}
and
\begin{equation}
\d V \, = 
V\, \int_0^{2\pi} d\s \, \, W^{-1} B_{\mu\nu} \, W\,   
{d x^{\mu}\o{d\s}} \, \d x^{\nu}
\lab{varvflat}
\end{equation}

We now want to study how $U$ changes as we vary the volume $\O$, but 
keeping 
the point $x_0$ fixed. So, from \rf{ueq1} one has
\begin{equation}
U^{-1}{d \d U \o {d \zeta}} + U^{-1}\,\d S(H,B,A, \ca ,\zeta ) \, U  
+ 
U^{-1}\, S(H,B,A, \ca ,\zeta ) \,\d U= 0
\lab{varueq1}
\end{equation}
Again from \rf{ueq1} one gets
\begin{equation}
{d U^{-1} \o {d \zeta}} - U^{-1}\, S(H,B,A, \ca ,\zeta )  = 0
\lab{uinveq}
\end{equation}
and so
\begin{equation}
{d \, \o{d \zeta}} \( U^{-1} \d U \) = - U^{-1} \d S(H,B,A, \ca  , 
\zeta ) U 
\end{equation}

When evaluating $\d S(H,B,A, \ca  , \zeta )$, we perform an 
integration by 
parts whenever we have a derivative of $\d x^{\mu}$ and use the fact 
that
\begin{equation}
\d x^{\mu}(\s =0 ) = \d x^{\mu}(\s =2\pi ) = 
\d x^{\mu}(\tau =0 ) = \d x^{\mu}(\tau =2\pi ) = \d x^{\mu}(\zeta = 0 
) =0
\lab{endpoints4d}
\end{equation}

Then, performing the calculation one gets (without keeping the border 
of
$\Omega$ fixed)
\br  
& &U^{-1} \d U (\zeta ) =  
- U^{-1} \( \int_{0}^{2\pi} d \tau \, V \,\( 
\int_{0}^{2\pi} d \sigma \, 
\cw^{-1}\, H_{\mu\nu\rho} \, \cw
 \,{d x^{\mu}\o{d\sigma}} 
{d x^{\nu}\o{d\tau }}  \d x^{\rho}\) \, V^{-1}\) \, U \nonu\\ 
&-& \int_{0}^{\zeta} d \zeta^{\pr} \, U^{-1} \,\( \int_{0}^{2\pi} 
d \tau \,  V \,\( \int_{0}^{2\pi} d \sigma \times \right. \right. 
\nonu\\
&\times& \left. \left.  \cw^{-1}\, \( \cd_{\l} H_{\mu\nu\rho} - 
\cd_{\mu} H_{\nu\rho\l} + 
\cd_{\nu} H_{\rho\l\mu} -  \cd_{\rho} H_{\l\mu\nu} \) \cw \,
{d x^{\mu}\o{d\sigma}}{d x^{\nu}\o{d\tau }} {d 
x^{\rho}\o{d\zeta^{\pr}}} 
\d x^{\l}\)\,  V^{-1}\right. \nonu\\ 
&-& \left. 
\lb S(H,B,A ,\ca , \zeta^{\pr} ) \, , \,  \int_{0}^{2\pi} 
d \tau V\(  \int_{0}^{2\pi}  d \sigma  \, 
 \cw^{-1}\, H_{\mu\nu\rho} \cw  \,
   {d x^{\mu}\o{d\sigma}} {d x^{\nu}\o{d\tau }} \d x^{\rho}\)V^{-1}
\rb \) \, U  
\nonu\\ 
&-&  \int_{0}^{\zeta} d \zeta^{\pr} \, U^{-1} \, \int_{0}^{2\pi} 
d \tau \, V \,  \( 
\lb T\( B,A, \d \)\, , \, 
\int_{0}^{2\pi} d \sigma \cw^{-1}\, H_{\mu\nu\rho}\, \cw  
{d x^{\mu}\o{d\sigma}} {d x^{\nu}\o{d\tau }} {d 
x^{\rho}\o{d\zeta^{\pr}}}
\rb \right. 
\nonu\\
&-& \left. 
\lb T\( B,A, {d\, \o d\tau } \) \, , \,
 \int_{0}^{2\pi} d \sigma \cw^{-1}\,H_{\mu\nu\rho} \, \cw  
{d x^{\mu}\o{d\sigma}} {d x^{\rho}\o{d\zeta^{\pr}}} \d x^{\nu}
\rb \right. \nonu\\
&-& \left. 
\lb T\( B,A, {d\, \o d\zeta^{\pr}} \)\, , \, 
 \int_{0}^{2\pi} d \sigma\cw^{-1}\, H_{\mu\nu\rho}\, \cw  
{d x^{\mu}\o{d\sigma}} {d x^{\nu}\o{d\tau }} \d x^{\rho}\rb \) V^{-1} 
U
\lab{resultu1}
\er
where we have denoted
\begin{equation}
T\( B,A, * \) \equiv \int_0^{2\pi} d\s W^{-1} B_{\mu\nu} W {d 
x^{\mu}\o{d\s}}  
* x^{\nu}
\end{equation}
and have introduced the covariant derivative
\begin{equation}
\cd_{\l} H_{\mu\nu\rho} \equiv \pa_{\l} H_{\mu\nu\rho} + 
\lb \ca_{\l} \, , \, H_{\mu\nu\rho}
\rb 
\end{equation}

If we perform the integration up to $\zeta = 2 \pi$, and keep the 
border 
$\pa\O$ fixed, i.e.
\begin{equation}
\d x^{\mu}(\zeta = 2 \pi ) = 0 
\end{equation}
we get that the first term on the r.h.s. of \rf{resultu1} vanishes. 

We now consider the case where the three dimensional volume $\O$ is 
closed, 
and so $\pa\O$ is reduced to a point. Let ${\cal M}$ be the 
four-volume which 
boundary is $\O$. We can now scan ${\cal M}$ with closed 
three-volumes which 
have the fixed point $x_0$ in common. We parametrize such closed 
three-volumes 
by $\xi$ such that $\xi =0$ correspond to the infinitesimal volume 
around 
$x_0$ and $\xi = 2 \pi$ to the border $\O$. We can then consider the 
variation 
discussed above to correspond to the deformation of one closed 
three-volume 
into the other, and we write $\d \equiv d\xi {d\,\o d\xi}$. Denoting 
by $U_c$ 
the quantity $U$ integrated over a closed three-volume we then obtain 
from 
\rf{resultu1} 
\begin{equation}
{d U_c \o {d \xi}} +  U_c\,\(\int_{0}^{2\pi} \, d\zeta U^{-1} {\cal 
C} U \) =0
\lab{ucloseeq}
\end{equation}  
where we have introduced the quantity
\br
{\cal C} &=& 
\int_{0}^{2\pi} 
d \tau \,  V \,\( \int_{0}^{2\pi} d \sigma \times \right.  \nonu\\
&\times&  \left.  \cw^{-1}\, \( \cd_{\l} H_{\mu\nu\rho} - 
\cd_{\mu} H_{\nu\rho\l} + 
\cd_{\nu} H_{\rho\l\mu} -  \cd_{\rho} H_{\l\mu\nu} \) \cw \,
{d x^{\mu}\o{d\sigma}}{d x^{\nu}\o{d\tau }} {d x^{\rho}\o{d\zeta}} 
{d x^{\l}\o {d\xi}} \right. \nonu\\ 
&-& \left.  
\left[ T\( B,A, {d\,\o{d\xi}} \)\, , \, 
\int_{0}^{2\pi} d \sigma \cw^{-1}\, H_{\mu\nu\rho}\, \cw  
{d x^{\mu}\o{d\sigma}} {d x^{\nu}\o{d\tau }} {d x^{\rho}\o{d\zeta}}
\right] \right. 
\nonu\\
&-& \left. 
\left[ T\( B,A, {d\, \o d\tau } \) \, , \,
 \int_{0}^{2\pi} d \sigma \cw^{-1}\,H_{\mu\nu\rho} \, \cw  
{d x^{\mu}\o{d\sigma}} {d x^{\rho}\o{d\zeta}} {d x^{\nu}\o{d\xi}}
\right] \right. \nonu\\
&-& \left. 
\left[ T\( B,A, {d\, \o d\zeta} \)\, , \, 
 \int_{0}^{2\pi} d \sigma\cw^{-1}\, H_{\mu\nu\rho}\, \cw  
{d x^{\mu}\o{d\sigma}} {d x^{\nu}\o{d\tau }} {d x^{\rho}\o{d\xi}}\right] 
\) V^{-1} 
\nonu\\
&-& 
\lb S(H,B,A ,\ca , \zeta ) \, , \,  S(H,B,A ,\ca , \xi ) \rb  
\lab{curv4d} 
\er

The fact that $U$ for closed three-volumes can be obtained by 
integrating 
either \rf{ueq1} or \rf{ucloseeq} allows us to formulate another 
generalized 
non abelian Stokes theorem. Integrating \rf{ueq1} over a closed 
three-volume 
$\O$ and \rf{ucloseeq} over a four-volume ${\cal M}$ which boundary 
is $\O$ 
one gets  
\begin{equation}
{\hat {\cal P}} \exp \( \int_{\O} d\zeta d\tau \, V \, \(  d\sigma \,
\cw^{-1}\, H_{\mu\nu\rho} \cw \, {dx^{\mu}\o{d\sigma}} 
{dx^{\nu}\o{d\tau}}{dx^{\rho}\o{d\zeta}}\) \, V^{-1}\) = 
{\tilde {\cal P}} \exp \( \int_{{\cal M}} d\xi d\zeta  \, U^{-1} 
{\cal C} U \)
\lab{stokes4d}
\end{equation}
where ${\hat {\cal P}}$ means three-volume ordering and ${\tilde 
{\cal P}}$ 
four-volume ordering. 

Notice from \rf{resultu1} and \rf{curv4d} that ${\cal C}=0$ is a 
sufficient 
condition for $U$ to be independent of the three-volume $\O$, as long 
as its 
border $\pa\O$ is kept fixed. In particular, if $\O$ is a closed 
volume it 
follows that $U$ integrated over $\O$ is equal to its initial value 
at 
$x_0$, i.e. $U(x_0)$, because due to volume independence we can 
continuously 
deform $\O$ to the infinitesimal volume around $x_0$. 

The construction of the conserved charges is a messy generalization of 
the discussion given in Section~\ref{sect:charges-3d}.  Assume 
spacetime is ${\bf R}\times S^3$.  We briefly state that if we take 
the three surface to be a fixed time three sphere and if the curvature 
\rf{curv4d} vanishes then characters of $U$ are constants of the 
motion.

\subsection{Local integrability conditions}

The construction of the conserved quantities discussed above requires 
the 
vanishing of the curvature ${\cal C}$ \rf{curv4d}. However, that is a 
highly 
non local quantity. We now discuss how to impose local conditions 
which 
guarantee the vanishing of ${\cal C}$. 

In the three dimensional case we have seen two manners of imposing 
local 
integrability conditions. In the four dimensional case the local 
conditions 
are very similar. 

We need in fact the vanishing of four curvatures, namely
\begin{equation}
F_{\mu\nu}=0 \; ; \qquad {\cal F}_{\mu\nu}=0 \; ; \qquad 
{\cal K} = 0 \; ; \qquad {\cal C} = 0
\lab{4zerocurv}
\end{equation}
The first two imply 
\begin{equation}
A_{\mu} = - \pa_{\mu}\, W \, W^{-1} \; \qquad 
\ca_{\mu} = - \pa_{\mu}\, \cw \, \cw^{-1}
\lab{flat2a}
\end{equation}

Now, the vanishing of ${\cal K}$, as we saw in subsection 
\ref{subsec::local3d}, can be achieved with two types of local 
conditions 

\begin{enumerate}
\item By choosing $B_{\mu\nu}$ on a given point of space time to be 
$B_{\mu\nu}^{(0)}$ such that 
\begin{equation}
B_{\mu\nu} (x) \equiv W(x) B_{\mu\nu}^{(0)} W^{-1}(x) \; ; \qquad 
\mbox{\rm with} \qquad 
\lb B_{\mu\nu}^{(0)} \, , \,  B_{\rho\s}^{(0)} \rb =0
\end{equation}
which implies 
\begin{equation}
D_{\l} B_{\mu\nu} = 0 
\end{equation}

\item By choosing
\begin{equation}
A_{\mu} = A_{\mu}^a T_a \; , \qquad B_{\mu\nu} = B_{\mu\nu}^i P_i
\end{equation}
with $T_a$ and $P_i$ satisfying \rf{rt}, and imposing 
\begin{equation} 
D_{\l} B_{\mu\nu} + D_{\mu} B_{\nu\l} +  D_{\nu} B_{\l\mu} = 0
\end{equation}
which in four dimensions can be written as
\begin{equation}
D_{\mu} {\tilde B}^{\mu\nu} =0 \; ; \qquad \mbox{\rm with} \qquad 
{\tilde B}^{\mu\nu} \equiv \h \epsilon^{\mu\nu\rho\s} B_{\rho\s}
\lab{dual4db}
\end{equation}
\end{enumerate}

These two cases imply that $V$ has the form of a product of  
exponentials of generators of an abelian algebra (generators of 
$B_{\mu\nu}^{(0)}$ in the first case, and  $P_i$'s in the second). 
One way 
of canceling the terms involving $T(B,A,*)$ in \rf{curv4d} is to 
impose that 
such abelian subalgebras (where $T(B,A,*)$ lives) should commute with 
the 
algebra where $\cw^{-1} H_{\mu\nu\rho}\cw$ lives. But, if that is so, 
we 
observe that $V$ commutes with $\cw^{-1} H_{\mu\nu\rho}\cw$, and 
therefore 
$S(H,B,A, \ca ,\zeta )$, defined in \rf{rdef1} lives on the same 
algebra as 
$\cw^{-1} H_{\mu\nu\rho}\cw$. So, one way of canceling the last term 
of 
\rf{curv4d} is to impose that the algebra where $\cw^{-1} 
H_{\mu\nu\rho}\cw$ 
lives is also abelian. Therefore, the vanishing of ${\cal C}$ 
requires now the 
local condition
\begin{equation}
 \cd_{\l} H_{\mu\nu\rho} - 
\cd_{\mu} H_{\nu\rho\l} + 
\cd_{\nu} H_{\rho\l\mu} -  \cd_{\rho} H_{\l\mu\nu} = 0
\end{equation}

Therefore, we shall consider four cases of local conditions for the 
vanishing 
of the curvatures \rf{4zerocurv}. In analogy with \rf{rt}, we 
introduce the 
algebra
\br
\lb T_a \, , \, T_b \rb &=& f_{ab}^c T_c \qquad \qquad \; \; \; 
\lb \ctt_r \, , \, \ctt_s \rb = C_{rs}^u \ctt_u\nonu\\
\lb T_a \, , \, P_i \rb &=& P_j R_{ji}\( T_a\) \qquad \, 
\lb \ctt_r \, , \, \cs_m \rb = \cs_n {\cal R}_{nm}\( \ctt_r\) \nonu\\
\lb P_i \, , \, P_j \rb &=& 0 \qquad \qquad \qquad  
\lb \cs_m \, , \, \cs_n \rb = 0
\lab{rtcal}
\er
and
\begin{equation}
\lb P_i \, , \, \cs_m \rb =0
\lab{con4d1b}
\end{equation}
where again, due to the Jacobi identities, $R$ and ${\cal R}$ 
constitute  
matrix representation of the algebra of $T_a$'s and $\ctt_r$'s 
respectively, 
i.e.
\begin{equation}
\lb R\( T_a \) \, , \, R\( T_b\)  \rb =  R\( \lb T_a \, , \, T_b \rb 
\) 
\qquad 
\lb {\cal R}\( \ctt_r\) \, , \,{\cal R}\( \ctt_s\) \rb = 
{\cal R}\( \lb \ctt_r \, , \, \ctt_s \rb \)
\end{equation}
Notice that we are not assuming anything about the commutation 
relations 
$\lb T_a \, , \, \ctt_r \rb$, $\lb T_a \, , \, \cs_m \rb$, and 
$\lb \ctt_r \, , \,  P_i \rb $. That is because they are not relevant 
for the 

The local conditions are
\begin{itemize}
\item {\bf Type I)} We take
\begin{equation}
A_{\mu} = A_{\mu}^a \, T_a \; ; \qquad \ca_{\mu} = \ca_{\mu}^r \, 
\ctt_r 
 \; ; \qquad B_{\mu\nu} = B_{\mu\nu}^i \, P_i  
 \; ; \qquad H_{\mu\nu\rho} =  H_{\mu\nu\rho}^m \, \cs_m
\lab{con4d1a}
\end{equation}
and also
\begin{equation}
F_{\mu\nu} = 0 \; \qquad \flie_{\mu\nu} = 0 \; \qquad 
D_{\mu} {\tilde B}^{\mu\nu} = 0 \; \qquad 
\cd_{\mu} {\tilde H}^{\mu} = 0
\lab{con4d1c}
\end{equation}
where 
\begin{equation}
{\tilde H}^{\mu}  \equiv {1\o 3} 
\epsilon^{\mu\nu\rho\sigma}H_{\nu\rho\sigma} 
\lab{con4d1d}
\end{equation}
and ${\tilde B}^{\mu\nu}$ is defined in \rf{dual4db}.

\item  {\bf Type II)} We take
\br
A_{\mu} = A_{\mu}^a \, T_a \; ; & \quad& B_{\mu\nu} = B_{\mu\nu}^i \, P_i
\nonu \\
H_{\mu\nu\rho}(x) \equiv \cw (x) H_{\mu\nu\rho}^{(0)} 
\cw^{-1}(x) 
\; ; & \quad& \lb  H_{\mu\nu\rho}^{(0)} \, , \, 
H_{\mu^{\pr}\nu^{\pr}\rho^{\pr}}^{(0)} \rb = 0 
\lab{con4d2a}
\er
with 
\begin{equation}
 \lb  H_{\mu\nu\rho}^{(0)} \, , \, P_i \rb = 0
\lab{con4d2b}
\end{equation}
and also
\begin{equation}
F_{\mu\nu} = 0 \; \qquad \flie_{\mu\nu} = 0 \; \qquad 
D_{\mu} {\tilde B}^{\mu\nu} = 0 
\lab{con4d2c}
\end{equation}
Notice that \rf{con4d2a} and \rf{con4d2c} imply
\begin{equation}
\cd_{\l} H_{\mu\nu\rho}=0
\lab{con4d2d}
\end{equation}

\item {\bf Type III)} We take
\begin{equation}
\ca_{\mu} = \ca_{\mu}^r \, \ctt_r \; \quad  
H_{\mu\nu\rho} =  H_{\mu\nu\rho}^m \, \cs_m\; \quad 
B_{\mu\nu}(x) = W(x) B_{\mu\nu}^{(0)} W^{-1}(x)  \; \quad 
\lb B_{\mu\nu}^{(0)} \, , \, B_{\rho\s}^{(0)}\rb = 0 
\lab{con4d3a}
\end{equation}
with
\begin{equation}
\lb B_{\mu\nu}^{(0)} \, , \, \cs_m \rb = 0 
\lab{con4d3b}
\end{equation}
and also
\begin{equation}
F_{\mu\nu} = 0 \; \quad \flie_{\mu\nu} = 0 \; \quad 
\cd_{\mu} {\tilde H}^{\mu} = 0  
\lab{con4d3c}
\end{equation}
Notice that \rf{con4d3a} and \rf{con4d3c} imply
\begin{equation}
D_{\rho} B_{\mu\nu} = 0 
\lab{con4d3d}
\end{equation}

\item {\bf Type IV)} Finally we take
\br
B_{\mu\nu}(x) &=& W(x) B_{\mu\nu}^{(0)} W^{-1}(x)  \; \quad 
 H_{\mu\nu\rho}(x) \equiv \cw (x) H_{\mu\nu\rho}^{(0)} \cw^{-1}(x) 
\nonu \\
\lb B_{\mu\nu}^{(0)} \, , \, B_{\rho\s}^{(0)}\rb &=& 0   \; \quad 
 \lb  H_{\mu\nu\rho}^{(0)} \, , \, 
H_{\mu^{\pr}\nu^{\pr}\rho^{\pr}}^{(0)} \rb = 0
\lab{con4d4a}
\er
with 
\begin{equation}
 \lb  H_{\mu\nu\rho}^{(0)} \, , \,  B_{\mu^{\pr}\nu^{\pr}}^{(0)} \rb 
= 0
\lab{con4d4b}
\end{equation}
and also
\begin{equation}
F_{\mu\nu} = 0 \; \qquad \flie_{\mu\nu} = 0 
\lab{con4d4c}
\end{equation}
Notice that \rf{con4d4a} and \rf{con4d4c} imply
\begin{equation}
D_{\rho} B_{\mu\nu} = 0 \; \qquad \cd_{\l} H_{\mu\nu\rho}=0 
\lab{con4d4d}
\end{equation}

\end{itemize}

In the cases of the integrability conditions of types {\bf I} and 
{\bf II} we 
have the conserved currents 
\begin{equation}
K^{\mu\nu} \equiv W^{-1} {\tilde B}^{\mu\nu} W \qquad \qquad 
\pa_{\mu} K^{\mu\nu} = 0 
\lab{tensorcons}
\end{equation}
and in the cases of types {\bf I} and {\bf III} we have 
\begin{equation}
{\cal J}^{\mu} \equiv \cw^{-1} {\tilde H}^{\mu} \cw 
\qquad \qquad  \pa_{\mu} {\cal J}^{\mu} = 0 
\lab{conscurcal}
\end{equation}

In all four cases we have the gauge symmetries
\br
A_{\mu} &\ra & g \, A_{\mu} \, g^{-1} - \pa_{\mu} g \, g^{-1} \nonu\\
B_{\mu\nu} &\ra & g \, B_{\mu\nu} \, g^{-1}  
\er
and
\br
\ca_{\mu} &\ra & G \, \ca_{\mu} \, G^{-1} - \pa_{\mu} G \, G^{-1} 
\nonu\\
H_{\mu\nu\rho} &\ra & G \, H_{\mu\nu\rho} \, G^{-1}  
\er
where $g$ and $G$, according to each case, are elements of groups 
whose Lie 
algebras are those of 
$A_{\mu}$ and $\ca_{\mu}$ respectively. The conserved currents ${\cal 
J}^{\mu}$ 
and $K^{\mu\nu}$ are invariant under these gauge transformations. 

In the cases of the integrability conditions of types  {\bf I} and 
{\bf II}  
we have the additional gauge symmetries
\br
A_{\mu} &\ra &  A_{\mu}  \nonu\\
B_{\mu\nu} &\ra & B_{\mu\nu} + D_{\mu} \a_{\nu} - D_{\nu} \a_{\mu}
\er
where the parameters $\a_{\mu}$ take values in the abelian algebra 
generated by 
the $P_i$'s. Under these transformations, the currents $K^{\mu\nu}$ 
transform 
as
\begin{equation}
K^{\mu\nu} \ra K^{\mu\nu} + \epsilon^{\mu\nu\rho\sigma}\, \pa_{\rho} 
\( W^{-1} \a_{\sigma} W \) 
\end{equation}

For the integrability conditions of types  {\bf I} and {\bf III} 
we have the gauge symmetries
\br
\ca_{\mu} &\ra &  \ca_{\mu}  \nonu\\
H_{\mu\nu\rho} &\ra & H_{\mu\nu\rho} + \cd_{\mu} \b_{\nu\rho} + 
\cd_{\nu} \b_{\rho\mu} + \cd_{\rho} \b_{\mu\nu} 
\lab{3gauge}
\er
where the parameters $\b_{\mu\nu}$ ($=-\b_{\nu\mu}$) take values in 
the abelian 
algebra generated by the $\cs_m $'s. The conserved currents ${\cal 
J}^{\mu}$ 
transform, under \rf{3gauge}, as 
\begin{equation}
{\cal J}^{\mu} \ra {\cal J}^{\mu} + \epsilon^{\mu\nu\rho\sigma}\, 
\pa_{\nu} 
\( \cw^{-1} \b_{\rho\sigma} \cw \) 
\lab{3gaugecurr}
\end{equation}

\sect{Curvature in loop space}
\label{sect:curv}

We now show that the explicit results of the previous sections are 
concrete realizations of a geometrical principle: the vanishing of
 curvature in an appropriate loop space.

\subsection{Review of principal bundles: general theory}

Assume we have a manifold $M$ and we would like to construct a principal 
bundle $P$ with structure group $G$.  The construction of $P$ proceeds 
as follows.  Let $\{U_\alpha\}$ be an open cover for $M$.  Locally the 
bundle $P$ is isomorphic to $U_\alpha\times G$.  The bundle is defined 
by specifying some transition functions $\phi_{\alpha\beta}: 
U_\alpha\cap U_\beta \to G$.  On a nonempty overlap $U_\alpha \cap 
U_\beta$ the point $(x,g_\alpha)\in U_\alpha\times G$ is identified 
with $(x,g_\beta)\in U_\beta\times G$ via $g_\alpha 
\phi_{\alpha\beta} = g_\beta$.  A connection on the bundle may be 
specified in two different but equivalent ways.  There is a local 
definition defined in terms of the cover.  A connection is given by a 
collection of Lie algebra valued one forms $\{A_\alpha\}$ such that on 
$U_\alpha \cup U_\beta$ one has $A_\beta = \phi^{-1}_{\alpha\beta} 
d\phi_{\alpha\beta} + \phi^{-1}_{\alpha\beta} A_\alpha 
\phi_{\alpha\beta}$.  The second definition is global and may be 
inferred from the local cover definition.  At $(x,g_\alpha) \in 
U_\alpha \times G$ consider the one form $\omega_\alpha = g_\alpha A_\alpha 
g^{-1}_\alpha - dg_\alpha g^{-1}_\alpha$. One can easily check that 
over $U_\alpha \cap U_\beta$ the one forms $\omega_\alpha$ and 
$\omega_\beta$ agree, i.e., $\omega_\alpha=\omega_\beta$. Therefore, 
there exists a globally defined one form $\omega$ such that its 
restriction is precisely $\omega_\alpha$. Note that if we cavalierly 
write $\omega = g A g^{-1} - dg g^{-1}$ then 
\begin{eqnarray*}
	d\omega & = &g (dA) g^{-1} + dg\wedge A g^{-1} +
	g A \wedge g^{-1} dg g^{-1} - dg g^{-1} \wedge dg g^{-1}  \\
	 & = & - \omega \wedge \omega + g(dA + A\wedge A) g^{-1}  \\
\end{eqnarray*}
Since $\omega$ is globally defined we have that $\Omega = d\omega + 
\omega\wedge\omega$ is also globally defined.  This is the global 
definition of the curvature and it is related to the local one by 
$\Omega = g F g^{-1}$.  For completeness we  give the intrinsic 
definition of $\omega$.  A connection $\omega$ is a Lie algebra valued 
$1$-form on $P$ such that its restriction to the fiber is the right 
invariant $1$-form $-dg g^{-1}$ on $G$, and such that under the left 
action of $G$ on $P$ by (a constant) element $h$ one has $\omega \to h 
\omega h^{-1}$.  

In addition we need the parallel transport operator $W$.  We need to 
be a bit more explicit with notation in this section thus we will 
briefly reiterate previous definitions with more attention to the 
parameter $\sigma$ along curve.  Let $x_0$ be a fixed point on $M$ and 
let $x(\sigma)$ be a curve then parallel transport from $x_0=x(0)$ to 
$x(\sigma)$ is the solution to the ordinary differential equation
$$
	\frac{d}{d\sigma} W(\sigma) + A_\mu(x(\sigma))\frac{dx^\mu}{d\sigma} 
	W(\sigma) = 0\;.
$$
with initial condition $W(0)=I$.

None of the above discussions depends on the manifold $M$ being 
finite dimensional and thus can be applied to the infinite dimensional 
case. This is what we will do.

\subsection{Connections on loop spaces}
\label{sub:conn}

Assume we have a principal bundle $P\to M$ with connection.  For a 
fixed point $x_0\in M$ let $\Omega(M,x_0)$ be the space of all loops 
based at $x_0$:
$$
	\Omega(M,x_0) = \{ \gamma: S^1 \to M \mid \gamma(0)=x_0\}\;.
$$
we now want to construct a principal $G$-bundle over $\Omega(M,x_0)$ 
with connection.  Note that the structure group of the bundle is a 
finite dimensional group not a loop group.  It will be the trivial 
bundle $\xxxcal{P}= \Omega(M,x_0)\times G$.  Conceptually the bundle 
is constructed as follows.  Over $x_0 \in M$ the bundle $P\to M$ has 
fiber $P_{x_0}$ which is isomorphic to $G$.  All loops in 
$\Omega(M,x_0)$ have $x_0\in M$ as a starting point so we can consider 
them having $P_{x_0}$ in common.  This is the common fiber in the 
cartesian product $\Omega(M,x_0)\times G$.  Mathematically we have a 
natural map $\pi:\Omega(M,x_0) \to M$ given by $\pi(\gamma)=x_0$.  The 
bundle $\xxxcal{P}$ is just the pullback bundle $\pi^* P$, see 
\ct{nepo}.  Since 
$\xxxcal{P} \to \Omega(M,x_0)$ is a trivial bundle we can put the 
trivial connection on it.  There is a more interesting connection one 
can put on it which exploits the connection on the bundle $P\to M$.  
Consider a Lie algebra valued $2$-form $B$ on $M$ such that under the 
transition function $\phi$ we have $B \to \phi^{-1}B\phi$.  Let 
$W(\sigma)$ be the parallel transport operator from the point 
$x(0)=x_0$ to the point $x(\sigma)$ along the loop $\gamma$.  We can 
assign the Lie algebra valued $1$-form
\begin{equation}
	\xxxcal{A}[x(\sigma)] = \int_0^{2\pi} d\sigma 
	\; W(\sigma)^{-1}B_{\mu\nu}(x(\sigma))W(\sigma)
	\frac{dx^\mu}{d\sigma}
	\delta x^\nu(\sigma)
	\label{calA}
\end{equation}
The transformation laws of the above are determined by $\phi(x_0)$ 
which is clearly associated with the common fiber $P_{x_0}$. Thus we can 
define a connection on $\xxxcal{P}$ by
$$
	\varpi = -dg g^{-1} + g\left( \int_0^{2\pi} d\sigma 
	\; W(\sigma)^{-1}B_{\mu\nu}(x(\sigma))W(\sigma)
	\frac{dx^\mu}{d\sigma}
	\delta x^\nu(\sigma)
		 \right) g^{-1} \;.
$$
Thus we can treat $\xxxcal{A}$ as the connection in a certain 
trivialization.  The curvature is given by $\xxxcal{F} = 
\delta\xxxcal{A} + \xxxcal{A}\wedge\xxxcal{A}$.  The curvature 
$\xxxcal{F}$ becomes expression~(\ref{eq:fdef}) when we specialize to a 
flat connection $A$.  

\begin{figure}[tbh]
	\centerline{\epsfxsize=0.5\textwidth\epsfbox{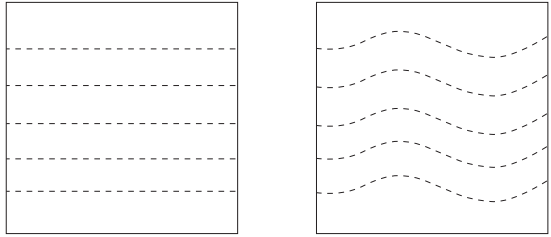}}
	\caption[squares]{\parbox[t]{.7\textwidth}{Decomposition ambiguity. 
	Dotted lines are constant $\tau$ curves.}}
	\protect\label{fig-squares}
\end{figure}

What do we mean when we say that we want to have parallel transport 
independent of path in $\Omega(M,x_0)$? Look at the space 
$\Omega(M,x_0)$ and consider a curve $\Gamma$ in $\Omega(M,x_0)$ 
parametrized by $\tau$ such that $\Gamma(0)$ is the ``constant curve'' 
$x_0$.  Note that for fixed $\tau$, $\Gamma(\tau)$ is a curve 
$x_\tau(\sigma)$ for $\sigma\in [0,2\pi]$ in $M$.  Thus it is convenient 
to ``write'' $\Gamma$ as $x(\sigma,\tau)$.  The statement that parallel 
transport be independent of the choice of curve $\Gamma \in 
\Omega(M,x_0)$ with fixed starting and ending points is the statement 
that the curvature vanish.
If one wants the parallel transport between points in $\Omega(M,x_0)$ 
to be independent of path then parallel transport should be path 
independent in $M$.  The reason for this is that a loop in 
$\Omega(M,x_0)$ beginning at the trivial loop may be viewed as a map 
from the square $[0,2\pi]^2$ to $M$ such that $\partial [0,2\pi]^2$ 
gets mapped to $x_0$.  Figure~\ref{fig-squares} shows two different 
sets of ``constant'' $\tau$ curves associated with the same closed 
$2$-submanifold in $M$.  To get the same result ordinary parallel 
transport should be path independent, i.e.  $F=0$.

\subsection{The curvature computation}
\label{sub:curva}

Below we explicitly describe the curvature computation. First we need 
the standard result:
\begin{eqnarray*}
	 W(\sigma)^{-1}\delta W(\sigma)& = & 
	 - W(\sigma)^{-1} A_\mu(x(\sigma)) 
	W(\sigma) \delta x^\mu(\sigma)  \\
	 & + & \int_0^\sigma d\sigma'\;
	W(\sigma')^{-1} F_{\mu\nu}(x(\sigma'))W(\sigma') \frac{dx^\mu}{d\sigma'}
	\delta x^\nu(\sigma')\;.
\end{eqnarray*}
We also need definition~(\ref{calA}).
Let $\delta$ be the exterior derivative on the space 
$\Omega(M,x_0)$ and thus $\delta^2=0$ and 
$$
	\delta x^\mu(\sigma)\wedge \delta x^\nu(\sigma') 
	=- \delta x^\nu(\sigma')\wedge\delta x^\mu(\sigma)\;.
$$

Computing the curvature $\xxxcal{F} = \delta\xxxcA + \xxxcA\wedge\xxxcA$ 
is  tedious, but straightforward, so we do not reproduce it in detail. 

It is convenient to introduce a notation which will simplify the 
formulas. For any object $X$ which transforms under the adjoint 
representation define the parallel transported object $X^W$ by
$$
	X^W(\sigma) = \wil^{-1} X(\sigma) \wil \;.
$$
An elementary exercise shows that
$$
	\frac{d}{d\sigma} X^W = \wil^{-1}(D_\mu X)\wil 
	\frac{dx^\mu}{d\sigma}\;.
$$

Using this notation we have, after partial integrations and use of antisymmetry,
we have:

\begin{eqnarray*}
	\delta\xxxcA 
	& = & - \int_0^{2\pi} d\sigma \int_0^\sigma d\sigma' \;
	 \left[ F^W_{\kappa\mu}(x(\sigma')) ,
	 B^W_{\lambda\nu}(x(\sigma)) \right] 
	\frac{dx^\kappa}{d\sigma'} \frac{dx^\lambda}{d\sigma}
	\delta x^\mu(\sigma') \wedge\delta x^\nu(\sigma) \\
	 & - &  \half\int_0^{2\pi} d\sigma \;
	 \wil^{-1} 
	 \left[ D_\lambda B_{\mu\nu}
	 + D_\mu B_{\nu\lambda}
	 + D_\nu B_{\lambda\mu}\right](x(\sigma)) 
	 \wil
	 \frac{dx^\lambda}{d\sigma}
	 \delta x^\mu(\sigma)\wedge \delta x^\nu(\sigma)
\end{eqnarray*}

To finish the curvature computation we need 
$$
\xxxcA\wedge\xxxcA = \half \int_0^{2\pi} d\sigma \int_0^{2\pi} d\sigma'
	\left[ B^W_{\kappa\mu}(x(\sigma')), B^W_{\lambda\nu}(x(\sigma))
	\right]
	\frac{dx^\kappa}{d\sigma'}\frac{dx^\lambda}{d\sigma}
	\delta x^\mu(\sigma') \wedge \delta x^\nu(\sigma)\;.
$$

Putting things together we have that the curvature $\xxxcal{F} = 
\delta\xxxcA + \xxxcA\wedge\xxxcA$ is given by
\begin{eqnarray*}
	\xxxcal{F}
	 & = & - \half\int_0^{2\pi} d\sigma \;
	 \wil^{-1} 
	 \left[ D_\lambda B_{\mu\nu}
	 + D_\mu B_{\nu\lambda}
	 + D_\nu B_{\lambda\mu}\right](x(\sigma)) 
	 \wil \\
	 & &  \mathstrut\hspace{.5in} \times
	 \frac{dx^\lambda}{d\sigma}
	 \delta x^\mu(\sigma)\wedge \delta x^\nu(\sigma) \\
	& - & \int_0^{2\pi} d\sigma \int_0^\sigma d\sigma' \;
	 \left[ F^W_{\kappa\mu}(x(\sigma')) ,
	 B^W_{\lambda\nu}(x(\sigma)) \right] 
	\frac{dx^\kappa}{d\sigma'} \frac{dx^\lambda}{d\sigma}
	\delta x^\mu(\sigma') \wedge\delta x^\nu(\sigma) \\
	& + &
	\half \int_0^{2\pi} d\sigma \int_0^{2\pi} d\sigma'
	\left[ B^W_{\kappa\mu}(x(\sigma')), B^W_{\lambda\nu}(x(\sigma))
	\right]
	\frac{dx^\kappa}{d\sigma'}\frac{dx^\lambda}{d\sigma}
	\delta x^\mu(\sigma') \wedge \delta x^\nu(\sigma)\;.
\end{eqnarray*}
With a slight change of notation this reduces to expression 
(\ref{eq:fdef}) when $F$ vanishes.

\subsection{Curvature in higher loop spaces}
\label{sect:higher-loop}

One can generalize the ideas in Section~\ref{sect:curv} to higher loop 
spaces.  There are at least two ways to generalize the method 
presented in three dimensions.  First there is the hierarchical 
approach we took to four dimension.  Here one first imposes zero 
curvature in the ordinary sense.  Then the loop space zero curvature 
condition on all the two dimensional surfaces and finally one 
deformation independence in three volumes.  In fact we slightly 
generalized things by introducing a second one form $\cal A$ which was 
used to do a different parallel transport on the three volumes.  It is 
clear that this hierarchical structure can be extended to arbitrary 
high dimensionality.  

In the examples we analyzed, we saw that only structure that entered 
in the four dimensional case was that associated with the three 
volumes and the associated parallel transport.  We could have saved a 
lot of work by just considering this simpler situation but we were 
curious about the more general case. If we assume that this is the 
generic structure then it is possible to derive the curvature of 
higher loop spaces. Proceeding as in the previous sections is 
algebraically very messy. It is worthwhile investing a little effort 
and doing things more mathematically and more abstractly. 

One defines the higher loop spaces inductively by 
$\Omega^{n+1}(M,x_0) = \Omega(\Omega^n(M,x_0),x_0)$. In layman's terms 
we have
\begin{eqnarray}
	\Omega^n(M,x_0) &=& \left\{ f: [0,1]^n \to M \bigg|
	f|_{\partial [0,1]^n} = x_0 \right\} 
	\label{the-square}\\
	&=& \left\{ f: S^n \to M \bigg|
	f(\mbox{north pole}) = x_0 \right\}
	\nonumber
\end{eqnarray}
Next we observe that a tangent vector $X$ at $N\in \Omega^n(M,x_0)$ is a 
vector field along $N$ (not necessarily tangential to $N$) 
representing how one deforms $N$.  Note that the vector field must 
vanish at $x_0$ because $x_0$ is kept fixed.  This replaces the role 
of $\delta x^\mu(\sigma)$ from our earlier discussions.

We now mimic what was done in Section~\ref{sub:conn}.  First we have a 
principal bundle $P\to M$ with structure group $G$ and connection 
$A$.  We assume from the beginning that $A$ is a flat connection 
because experience has taught us that this is necessary to avoid 
nonlocal behavior.  Next we construct the trivial principal bundle 
$\xxxcal{P} = \Omega^n(M,x_0) \times G$.  The fiber of this bundle is 
identified with $P_{x_0}$ as before.  We now introduce an $(n+1)$-form 
$B$ that transforms under the adjoint representation of the group $G$.  
Finally we construct a Lie algebra valued $1$-form which will be our 
connection on $\cal P$.  One forms are characterized by their values 
when applied to a vector.  Given a tangent vector $X$ at $N\in 
\Omega(M,x_0)$ the value of the connection is
\begin{equation}
	\xxxcA(X) = \int_N W^{-1}(\iota_X  B) W\;,
	\label{def:conn}
\end{equation}
where $\iota_X$ is interior multiplication by $X$.  Note that $\iota_X 
B$ is an $n$-form and thus suitable for integration over $N$.  In the 
above $W$ is parallel transport from $x_0$ to the relevant point.  
We remind the reader that the flatness of the connection ensures 
that $W$ is a local function of the endpoint. For convenience we define
$$
	B^W = W^{-1} B W
$$
then $B^W$ is only a function of the point\footnote{The basepoint 
$x_0$ is kept fixed and is thus omitted from our discussions.} $x \in 
M$.  To compute the curvature we need to take the 
exterior derivative.  We apply the standard formula of the exterior 
derivative of a $1$-form to $\xxxcA$ and obtain
\begin{equation}
	d\xxxcA(X,Y) = X\xxxcA(Y) - Y\xxxcA(X)
	- \xxxcA([X,Y])
	\label{extd}
\end{equation}
where $X$ and $Y$ are tangent vector fields in a neighborhood of  $N$. 
How do we interpret $X\xxxcA(Y)$, the directional derivative of the 
function $\xxxcA(Y)$ in direction $X$? We observe that $X$ generates 
a deformation of $N$ thus 
$$
	X\xxxcA(Y) = \int_N {\cal L}_X( \iota_Y B^W)\;,
$$
where ${\cal L}_X$ is the Lie derivative on $M$ along the flow $X$. 
Had we assumed that $A$ was not flat then we would have had to work 
much harder as in Section~\ref{sub:curva} because the action of $X$ 
on $W$ is nonlocal. Inserting into (\ref{extd}) we conclude that
\begin{equation}
	d\xxxcA(X,Y) = \int_N \left( {\cal L}_X \iota_Y B^W - 
	{\cal L}_Y \iota_X B^W - \iota_{[X,Y]} B^W \right)
	\label{extd1}
\end{equation}
We now make an observation which will allow us to compute the 
curvature in arbitrary dimension.  Let $\omega$ be a differential 
form and let $X$ and $Y$ be vector fields and define an operator 
$Q(X,Y)$ on forms by
\begin{equation}
Q(X,Y)\omega = {\cal L}_X \iota_Y \omega - 
	{\cal L}_Y \iota_X \omega - \iota_{[X,Y]}\omega
	- \iota_Y \iota_X d\omega + d(\iota_Y\iota_X \omega)
	\label{Qdef}
\end{equation}
We now enumerate properties of $Q$.
\begin{enumerate}
	\item  $Q(X,Y) = - Q(Y,X)$

	\item  $Q(X+Z,Y) = Q(X,Y) + Q(Z,Y)$ for vector fields $X,Y,Z$

	\item  $Q(X,Y)(\omega_1 + \omega_2) = Q(X,Y)\omega_1 + 
	Q(X,Y)\omega_2$ for forms $\omega_1,\omega_2$

	\item  $Q(fX,Y)\omega = f Q(X,Y)\omega$ for any function $f$

	\item  $Q(X,Y)(f\omega) = f Q(X,Y)\omega$ for any function $f$
\end{enumerate}
These properties are straightforward to prove and the only tricky 
identity required is ${\cal L}_{fX}\omega = f{\cal L}_X \omega + df 
\wedge \iota_X\omega$.  The above properties ensure that $Q$ is linear 
in all possible ways and thus the derivative terms all cancel.  
$Q(X,Y)$ is a linear operator and not a differential operator.  This 
is very analogous to curvature operators in geometry.  Because of 
linearity, all we need to determine $Q$ is its components with respect 
to any basis.  On $M$ choose $X = \partial/\partial x^\mu$, 
$Y=\partial/\partial x^\nu$ and $\omega = dx^{\lambda_1}\wedge 
dx^{\lambda_2}\wedge \cdots \wedge dx^{\lambda_p}$.  By inspection one 
sees that each individual term in (\ref{Qdef}) vanishes.  Since all 
the components vanish so must $Q$.

The vanishing of $Q$ tells us that (\ref{extd1}) may be written as
$$
	d\xxxcA(X,Y) = \int_N \left(\iota_Y\iota_X dB^W- d(\iota_Y\iota_X  
	B^W)\right)
$$
We have to be a little careful.  The parallel transport operator 
$W(x)$ from $x_0$ to $x$ is not only a function of the endpoints only but 
also of the homotopy class of the path.  One may potentially get a 
boundary contribution if $W$ is multiply valued.  In 
this case one can think of $N$ as defined by (\ref{the-square}).  The 
boundary contribution is not there because $X$ and $Y$ vanish at $x_0$.

The exterior derivative of $\xxxcA$ may be written as
\begin{eqnarray}
	d\xxxcA(X,Y) & = & \int_N \left(\iota_Y\iota_X dB^W\right)
	\label{dA-1}  \\
	 & = & \int_N W^{-1}\left(\iota_Y\iota_X DB \right)W
	\label{dA-2}
\end{eqnarray}
In (\ref{dA-2}) we have rewritten things in terms of the covariant 
derivative.  In computing the curvature ${\cal F} = d{\cal A} + {\cal 
A}\wedge{\cal A}$ we see that we will get a nonlocal term from the 
quadratic term in $\cal A$.  We postulate that integrability in higher 
dimension corresponds to a flat connections $A$ and $\cal A$.  
Requiring that ${\cal F}=0$  locally  requires that
\begin{equation}
	DB=0
	\label{hi-int1}
\end{equation}
and the vanishing of the the commutator term
\begin{equation}
	\left[\int_N \iota_X B^W, \int_N \iota_Y B^W\right]=0\;
	\label{hi-int2}
\end{equation}
for all vector fields $X$ and $Y$.
This is our generalization of integrability to higher  
dimension.

\sect{Examples}
\label{sect:Ex}
We work out  here some examples of models to show how 
they are easily fitted in out approach, and their integrability
can be therefore be explored  by our methods. We
present in more detail the simplest nontrivial case, which already has
interesting and encouraging new results. Others will follow in a future
publication.

\subsection{The example of the $CP^1$ model}
\label{sec:cp1}

We now show that the classical equations of motion of the $CP^1$ model in
$2+1$ dimensions can be written as \rf{wayout}, and therefore constitute
sufficient conditions for the surface independence of $V$, introduced in
\rf{veq}.

The $CP^1$ model has one complex scalar field $u$ and its equation of
motion is (see for instance \ct{ward})
\begin{equation}
\( 1 + \mid u \mid^2 \) \pa^2 u = 2 u^* \( \pa_{\mu} u\)\( \pa^{\mu} u\)
\lab{cp1}
\end{equation}
where $\pa^2 = \pa_{\mu}\pa^{\mu}= \pa_0^2 - \pa_1^2 - \pa_2^2$.

It can be obtained from the Lagrangian
\begin{equation}
{\cal L} = {\pa_{\mu}u\pa^{\mu}u^*\o{\( 1 + \mid u \mid^2\)^2}}
\end{equation}

The $CP^1$ model is invariant under the action of $SO(3)$. Associated
with the isometry
\begin{equation}
u \ra e^{i \theta} \, u
\lab{phase}
\end{equation}
one has Noether current
\begin{equation}
J_{\mu}^{\rm Noet.} = {1\o {\( 1 + \mid u \mid^2 \)^2 }} \( u^* \pa_{\mu} u -
u \pa_{\mu} u^* \)
\lab{noethercur}
\end{equation}
There are two more conserved Noether currents coming from the remaining 
isometries.  They are the real and imaginary parts of the current
\begin{equation}
j_{\mu} =   { {\pa_{\mu}u + u^2 \pa_{\mu} u^*} \o {\( 1 + \mid u\mid^2\)^2}}
\lab{isomcur}
\end{equation}
This model also has a topological charge given by\footnote{In fact, any
current of the form $\epsilon_{\mu\nu\l}f(u,u^*)\pa^{\nu} u^* \pa^{\l} u$, with
$ f(u,u^*)$ an arbitrary function of $u$ and $u^*$, is trivially
conserved. Any such current arises as the pullback of a $2$-form on
$CP^1$ and is thus manifestly closed. The cohomology class
$H^2(CP^1,{\bf R})$ is one dimensional so there is only one
independent conserved charge of this type.}
\begin{equation}
J^{\rm Top.}_{\mu} \equiv - {i\o 2\pi} \epsilon_{\mu\nu\l}
{{\pa^{\nu} u^* \pa^{\l} u}\o{\( 1+\u2 \)^2}}
\end{equation}
Notice that the Lagrangian and equations of motion of the $CP^1$ model are
invariant under the discrete symmetry
\begin{equation}
u \ra {1 \o u}
\lab{invsym}
\end{equation}

The $CP^1$ model is the same as the $O(3)$ sigma model which is
formulated in terms of three real scalar fields
$\bfphi = ( \phi_1,\phi_2,\phi_3)$ \ct{bp,wz}. It is defined by the
Lagrangian and constraint
\begin{equation}
{\cal L} = {1\o 2 g} \, (\pa_{\mu}\, \bfphi )^2 \qquad \qquad \bfphi^2 = 1
\lab{o3lagra}
\end{equation}
By adding to the Lagrangian a Lagrange multiplier term $\l ( \bfphi^2 -
1)$, one
gets the equations of motion to be
\begin{equation}
\pa^2 \bfphi + (\pa_{\mu}\,\bfphi \cdot \pa^{\mu}\,\bfphi ) \bfphi = 0
\lab{o3eq}
\end{equation}

The $O(3)$ sigma model is invariant under the global $SO(3)$ transformations
$\bfphi \ra M \cdot \bfphi$, with $M$ being orthogonal $3\times 3$ matrices.
The corresponding Noether charges are (write $M = \exp\( i \omega_j T_j\)$,
with $\lb T_i \, ,\, T_j \rb = i \epsilon_{ijk} T_k$)
\begin{equation}
j_{\mu}^{i} \equiv \epsilon^{ijk} \phi^{j}\pa_{\mu} \phi^{k} \, \qquad \qquad
i,j,k=1,2,3
\lab{so3cur}
\end{equation}
Notice, however, that such currents are constrained by
\begin{equation}
\sum_{i=1}^3\, \phi^{i} j_{\mu}^{i} =0
\lab{constrso3cur}
\end{equation}

The relation between the two models is achieved by the
stereographic projection of the $\bfphi$ field on the sphere to the $u$ field
on the plane. One has
\begin{equation}
u \equiv u_1 + i u_2 = {\phi_1 + i \phi_2 \o {1 - \phi_3}}
\lab{udef}
\end{equation}
or
\begin{equation}
\bfphi = {1\o {1+\mid u\mid^2}} \( u + u^* , \, -i(u - u^*) , \, \u2 -1 \)
\lab{phidef}
\end{equation}
Substituting \rf{udef} into \rf{cp1} one can verify that $CP^1$ is 
indeed equivalent to \rf{o3eq} after both are $\sigma$-models of the 
two sphere in the round metric.

These models have  Bogomolny type bounds which leads to first order
differential equations. Belavin and Polyakov \ct{bp} have shown that the
energy functional
\begin{equation}
E = \int d^2 x \,\(\(\pa_0 \bfphi\)^2 +  \(\pa_m \bfphi\)^2\)
\qquad m=1,2
\end{equation}
satisfies the bound $E \geq 8 \pi \, Q$, where $Q$ is the topological charge
\begin{equation}
Q = {1\o 8 \pi} \int d^2x \,
\epsilon_{ijk} \epsilon_{mn} \phi_i \pa_m \phi_j \pa_n \phi_k
\lab{topol}
\end{equation}
with $i,j,k=1,2,3$, and $m,n=1,2$.
The conditions for saturating the bound are
\br
\pa_0  \phi_i &=& 0 \nonu\\
\pa_m  \phi_i &=& \epsilon_{ijk} \epsilon_{mn} \phi_j \pa_n \phi_k
\lab{bpeqs}
\er
Any solution of these first order differential equations are solutions of the
$O(3)$ sigma model. In terms of the (static) $u$ fields,  eqs. \rf{bpeqs}
are the Cauchy-Riemann equations
\begin{equation}
\pa_{x_1}\, u_1 = \pa_{x_2}\, u_2 \, \qquad
\pa_{x_2}\, u_1 = -\pa_{x_1}\, u_2
\lab{creqs}
\end{equation}
and therefore any meromorphic function leads to a static solution of the
$CP^1$
model.

Of particular interest is the baby-Skyrmion solution which has one unit of
topological charge and corresponds to
\begin{equation}
u = x_1 + i x_2
\lab{bsusol}
\end{equation}
or
\begin{equation}
\bfphi = \( {x_1 \o r} \sin f(r) , \, {x_2 \o r} \sin f(r) , \, \cos f(r)\)
\, ; \qquad r^2\equiv x_1^2 + x_2^2
\lab{ansatz}
\end{equation}
with
\begin{equation}
f= 2\, \, {\rm arctg} {1\o r}
\end{equation}

We now show that the $CP^1$ model constitutes an example of the local
integrability conditions \rf{wayout}. We introduce the potentials
\br
A_{\mu} &=& {1\o{\( 1+\mid u \mid^2\) }} \(  -i \pa_{\mu} u \, T_{+}
-i \pa_{\mu} u^* \, T_{-} +  \( u \pa_{\mu} u^* - u^* \pa_{\mu} u \) \, T_3 \)
\lab{cp1pota}\\
{\tilde B}_{\mu} &=& {1\o{\( 1+\mid u \mid^2\)}}
 \( \pa_{\mu} u \, P_{1}^{(1)} - \pa_{\mu} u^* \, P_{-1}^{(1)} \)
\lab{cp1potb}
\er
where, following \rf{rt},  $T_3$, $T_{\pm}$ are the generators of the $sl(2)$
algebra
\br
\lb T_3 \, , \, T_{\pm} \rb &=& \pm  \, T_{\pm} \; , \qquad
\lb T_{+} \, , \, T_{-} \rb =  2\, T_3
\lab{sl2alg}
\er
and $P_{1}^{(1)}$ and  $P_{-1}^{(1)}$ (together with $P_{0}^{(1)}$) transform
under the  triplet
representation of it. For present and for future use, the commutation
relations
associated to a generic spin-$j$ representation of $sl(2)$
($m=-j,-j+1, \ldots , j-1, j$) are given by
\br
\lb T_3\, , \, P_{m}^{(j)} \rb &=&  m\,  P_{m}^{(j)}
\lab{sl2pa}\\
\lb T_{\pm}\, , \, P_{m}^{(j)} \rb &=&
\sqrt{j(j+1) - m(m\pm 1)} \,\,\, P_{m\pm 1}^{(j)}
\lab{sl2pb}\\
\lb P_{m}^{(j)}\, , \, P_{m^{\pr}}^{(j^{\pr})} \rb &=& 0
\lab{sl2pc}
\er

One can easily check that the gauge potential \rf{cp1pota} is flat
independent of the equations of motion of the $CP^1$ model
\begin{equation}
F_{\mu\nu} =0 \, ; \qquad  \qquad \mbox{\rm off shell.}
\lab{wayoutcp1a}
\end{equation}
Indeed, \rf{cp1pota} can be written as in \rf{flatamuw} with many
choices of $W$'s.  Two particularly useful choice are\footnote{Notice
that the commutation relations \rf{sl2alg}-\rf{sl2pc} are compatible
with the hermiticity conditions,
$T_3^{\dagger} = T_3$, $T_{\pm}^{\dagger} = T_{\mp}$ and
${P_{m}^{(j)}}^{\dagger}= (-1)^m P_{-m}^{(j)}$. Therefore,
$A_{\mu}^{\dagger} =- A_{\mu}$,
${\tilde B}_{\mu}^{\dagger} = {\tilde B}_{\mu}$, and
$W_1^{\dagger} = W_2^{-1}$. This guarantees that $J_{\mu}$ and
$J_{\mu}^{(j)}$, defined in \rf{conscurp} and \rf{conscurpj} respectively, are
hermitian since $J_{\mu}^{(j)} = W^{-1}_i \, {\tilde B}_{\mu}^{(j)} \, W_i$,
$i=1,2$, and
${J_{\mu}^{(j)}}^{\dagger} = W_1^{\dagger} \, {\tilde B}_{\mu}^{(j)} \,
{W^{-1}_1}^{\dagger} =
W^{-1}_2 \, {\tilde B}_{\mu}^{(j)} \, W_2 = J_{\mu}^{(j)}$.
Notice that, $W_1$ and $W_2$ are elements of the
group $SL(2, \IC )$, and not of $SU(2)$, but $i A_{\mu}$ does belong to the
algebra of $SU(2)$.
In addition, the mapping $\s (T_3) =-T_3$, $\s (T_{\pm}) = - T_{\mp}$, and
$\s (P_{m}^{(j)}) = (-1)^m P_{-m}^{(j)}$, is an automorphism of order $2$
of the algebra \rf{sl2alg}-\rf{sl2pc}.}
\begin{equation}
W= W_1 \equiv  e^{iuT_+}\, e^{\varphi T_3}\, e^{iu^*T_-} \, ; \qquad
\mbox{\rm and} \qquad
W = W_2 \equiv e^{iu^*T_-}\, e^{-\varphi T_3}\, e^{iuT_+}
\lab{w1w2}
\end{equation}
where $\varphi =  \ln (1 + \u2 )$.

On the other hand one can check that, as a consequence of the $CP^1$
equations of motion,
\begin{equation}
D_{\mu} {\tilde B}^{\mu} =0 \; , \qquad \qquad \mbox{\rm on shell.}
\lab{wayoutcp1b}
\end{equation}
In fact, \rf{wayoutcp1b} is equivalent to the $CP^1$ equations of motion.

The conserved currents \rf{invcur} in this case are (using $W=W_1$ or $W=W_2$)
\begin{equation}
J_{\mu} = W^{-1} \, {\tilde B}_{\mu} \, W =
 j_{\mu}\, P_{1}^{(1)} - i\sqrt{2}\, J_{\mu}^{\rm Noet.}\, P_{0}^{(1)} -
 j_{\mu}^* \, P_{-1}^{(1)}
\lab{conscurp}
\end{equation}
where $J_{\mu}^{\rm Noet.}$ and $j_{\mu}$ are respectively given by
\rf{noethercur} and \rf{isomcur}.

\subsubsection{A simpler submodel of $CP^{1}$}
\label{subsub:submodel}

In obtaining \rf{wayoutcp1b} it was important that $P_{1}^{(1)}$ and
$P_{-1}^{(1)}$ were respectively the highest and the lowest spin
states of the representation. If we now 
release that  restriction  by introducing
\begin{equation}
{\tilde B}_{\mu}^{(j)} = {1\o{\( 1+\mid u \mid^2\)}}
 \( \pa_{\mu} u \, P_{1}^{(j)} - \pa_{\mu} u^* \, P_{-1}^{(j)} \) \, \, ;
\qquad \mbox{\rm with $j$ integer and $j \geq 1$}
\lab{bfirstsub}
\end{equation}
one can easily verify that $D_\mu {\tilde B}_{\mu}^{(j)}=0$, 
with $A_{\mu}$ being the same as in \rf{cp1pota}):
\begin{eqnarray*}
	0 & = & \frac{1}{\left(1 + |u|^2\right)^2}\;\bigg\{
	\sqrt{j(j+1)-2} \left(
	-i  \partial_\mu u \partial^\mu u P^{(j)}_{+2}
	+i  \partial_\mu \bar{u} \partial^\mu \bar{u}
	 	 P^{(j)}_{-2} \right) \\
	 & + & \left((1+ |u|^2) \partial^2 u
	 - 2\bar{u} \partial_\mu u \partial^\mu u\right) P^{(j)}_{+1}
	  +  \left((1+ |u|^2) \partial^2 \bar{u}
	 - 2u \partial_\mu \bar{u} \partial^\mu \bar{u}\right) P^{(j)}_{-1}
	 \bigg\}
\end{eqnarray*}
As previously discussed, the $j=1$ case leads to the $CP^1$ equations
of motion. On the other hand, for all $j>1$, we see that
$D^{\mu} {\tilde B}_{\mu}^{(j)} =0$
{\it are} the equations of motion of the submodel of $CP^{1}$ \ct{ward}
\begin{equation}
\pa^2 \, u =0 \, ; \qquad \qquad \pa_{\mu} u \, \pa^{\mu} u =0
\lab{firstsub}
\end{equation}
Any solution of the above is a solution of the $CP^1$ model.
Conversely, one can easily verify that any (static) solution of the
Cauchy-Riemann equations or the Belavin-Polyakov equations \rf{creqs}
is a static solution of \rf{firstsub}.  Indeed, the baby-Skyrmion
\rf{bsusol} solves \rf{firstsub}.

So, from \rf{invcur} we expect $2j+1$ conserved currents. Since we are free
to choose $j$ we get in fact an infinite number of conserved currents.  Denote
\begin{equation}
J_{\mu}^{(j)} = W^{-1} \, {\tilde B}_{\mu}^{(j)} \, W \equiv
\sum_{m=-j}^{j} \, J_{\mu}^{(j,m)} \, P_{m}^{(j)}
\lab{conscurpj}
\end{equation}
The explicit new infinite currents are therefore given just by the 
elements $\ca_{m,\pm 1}$ of the adjoint action $W P^{(j)}_n 
W^{-1}=\sum^j_{m=-j}\ca_{mn}P^{(j)}_m$, which can be found in any book 
\ct{Gilmore}:
\begin{equation}
J^{(j,m)}_\mu ={{\pa_{\mu} u \ca_{m1} -  
\pa_{\mu} u^* \ca_{m -1}}\over {(1+{\mid u\mid}^2)}}
\lab{jexpl}
\end{equation}
The expressions are rather long and we give here instead the first 
three.  For the case $j=1$, we obviously get the same as in 
\rf{conscurp}.
For the case $j=2$ we get
\br
J_{\mu}^{(2,2)} &=&
-{{2\,i\,u \,\left( \pa_{\mu} u  +
       {{u }^2}\,\pa_{\mu} u^*  \right) }\over
   {{{\left( 1 + \mid u \mid^2  \right) }^3}}} \\
J_{\mu}^{(2,1)} &=&
{{(1  - 3\,\mid u \mid^2 ) \,\pa_{\mu} u  +
     (3  - \mid u \mid^2 ){{u }^2} \,\pa_{\mu} u^* }\over
   {{{\left( 1 + \mid u \mid^2   \right) }^3}}} \\
J_{\mu}^{(2,0)} &=&
-{{i\,{\sqrt{6}}\,\left( 1 - \mid u \mid^2   \right) \,
     \left(  u^* \,\pa_{\mu} u    -
       u \,\pa_{\mu} u^*  \right) }\over
   {{{\left( 1 + \mid u \mid^2  \right) }^3}}}
\er
with $J_{\mu}^{(2,-m)} = (-1)^m {J_{\mu}^{(2,m)}}^{\dagger}$, $m=1,2$.
For the case $j=3$ we have
\br
J_{\mu}^{(3,3)} &=&
-{{{\sqrt{15}}\,{{u }^2}\,
      \left( \pa_{\mu} u  + {{u }^2}\,\pa_{\mu} u^*  \right) }\over
    {{{\left( 1 + \mid u \mid^2  \right) }^4}}}\\
J_{\mu}^{(3,2)} &=&
{{i\,{\sqrt{10}}\,u \,
     \left( (-1  + 2  \mid u \mid^2 ) \,\pa_{\mu} u  +
(- 2  +
       \mid u \mid^2 )\,{{u }^2}\,\pa_{\mu} u^* \right) }\over
   {{{\left( 1 +  \mid u \mid^2   \right) }^4}}}\\
J_{\mu}^{(3,1)} &=&
{{( 1  - 8  \mid u \mid^2  +
     6  \mid u \mid^4) \,\pa_{\mu} u  +
     ( 6   -
     8  \mid u \mid^2  +
      \mid u \mid^4 ){{u }^2}\,\pa_{\mu} u^* }\over
   {{{\left( 1 +  \mid u \mid^2   \right) }^4}}}\\
J_{\mu}^{(3,0)} &=&
-{{2\,i\,{\sqrt{3}}\,\left( 1 - 3 \mid u \mid^2  +
       \mid u \mid^4 \right) \,
     \left(  u^* \,\pa_{\mu} u    -
       u \,\pa_{\mu} u^*  \right) }\over
   {{{\left( 1 +  \mid u \mid^2   \right) }^4}}}
\er
with $J_{\mu}^{(3,-m)} = (-1)^m {J_{\mu}^{(3,m)}}^{\dagger}$, $m=1,2,3$.

The submodel \rf{firstsub} has another representation for the integrability
conditions \rf{wayout}.
Let $P$, $Q$ and $\one$ be the generators of the Heisenberg algebra
\begin{equation}
\lb P \, , \, Q\rb =\one
\end{equation}
Notice that, within the scheme of \rf{rt} one can take $\cg = \{ Q \}$ and
$R =\{ P \, , \, \one \}$. Define
\begin{equation}
A_{\mu} = g(u) \pa_{\mu} u \, Q \; , \qquad
{\tilde B}_{\mu} = f(u) \pa_{\mu} u \, P
\end{equation}
where $f$ and $g$ are arbitrary nonvanishing functions of $u$. One can easily
verify that the connection $A_{\mu}$ is flat, i.e. $F_{\mu\nu}=0$, and that
\begin{equation}
D_{\mu} {\tilde B}^{\mu} = \( {d f\o{du}} \pa_{\mu} u \pa^{\mu} u + f \pa^2
u \) P
-  f g \pa_{\mu} u \pa^{\mu} u \, \one
\end{equation}
Therefore, the solutions of \rf{firstsub} solve the equations \rf{wayout}.

The conserved currents obtained from such representation are of the form
\begin{equation}
J_{\mu} = G(u) \pa_{\mu} u
\lab{generalcur}
\end{equation}
where $G(u)$ is any function of $u$ (and not of $u^*$). One has
$\pa_{\mu} J^{\mu}=0$ as a
consequence of the equations of motion \rf{firstsub}. Clearly, the complex
conjugate of \rf{generalcur} provide another set of conserved currents.

\subsubsection{Another submodel of $CP^1$}

Let us introduce the submodel of $CP^1$ defined by the equations
\begin{equation}
\( 1 + \mid u \mid^2 \) \pa^2 u = 2 u^* \( \pa_{\mu} u\)\( \pa^{\mu} u\)
\, \, ; \qquad \pa^{\mu} u \, \pa_{\mu} u^* = 0
\lab{secondsub}
\end{equation}
Notice that there are common solutions of \rf{firstsub} and \rf{secondsub}.
Solutions of the type $u=x_0 \pm x_j$ or
$u=i(x_0 \pm x_j)$, $j=1,2$, are common, but the baby-Skyrmion
$u = x_1 + i x_2$ does not solve \rf{secondsub}.

The submodel \rf{secondsub} admits some representations in terms of the
integrability conditions \rf{wayout}. Introduce the potentials
\begin{equation}
{\tilde B}_{\mu}^{(j,+)} = {1\o{\( 1+\mid u \mid^2\)}}
 \sum_{m=1}^j c(j,m) \pa_{\mu} u^m \, P_{m}^{(j)}  \, \, ;
\quad
{\tilde B}_{\mu}^{(j,-)} = {1\o{\( 1+\mid u \mid^2\)}}
 \sum_{m=1}^j c(j,m) \pa_{\mu} {u^*}^m \, P_{-m}^{(j)}
\lab{pmbj}
\end{equation}
with $j$ integer and $j \geq 1$, and where the coefficients $c(j,m)$ are
obtained, in terms of $c(j,1)$, from the recurrence relation
\begin{equation}
c(j,m) = {i\o m} \sqrt{j(j+1) - m(m-1)}\, \, c(j,m-1) \, \, ; \qquad \qquad
m=2,3,\ldots j
\end{equation}

One can check that, as a consequence of \rf{secondsub},
\begin{equation}
D^{\mu} \, {\tilde B}_{\mu}^{(j,\pm )} = 0
\end{equation}
with $A_{\mu}$ being the same as in \rf{cp1pota}.

Using the symmetry \rf{invsym}, one can observe that \rf{secondsub} admits the
following additional representations in terms of the integrability
conditions \rf{wayout}\footnote{One could use the same transformation in
\rf{cp1pota}-\rf{cp1potb} and \rf{bfirstsub}. However, unlike the present case,
no additional currents would be
obtained with that new representation of the integrability conditions.}
\br
A_{\mu}^{\rm inv}&=& {1\o{\( 1+\mid u \mid^2\) }} \(  i u^* \pa_{\mu} \log u
\, T_{+}
+i u \pa_{\mu} \log u^* \, T_{-} +   \pa_{\mu} \log ( {u\o u^*})  \, T_3 \)
\lab{potainv}\\
{\tilde B}_{\mu}^{(j,+),{\rm inv}} &=& {\mid u \mid^2\o{\( 1+\mid u \mid^2\)}}
 \sum_{m=1}^j c(j,m) \pa_{\mu} u^{-m} \, P_{m}^{(j)}
\lab{pmbjinv1} \\
{\tilde B}_{\mu}^{(j,-),{\rm inv}} &=& {\mid u \mid^2\o{\( 1+\mid u \mid^2\)}}
 \sum_{m=1}^j c(j,m) \pa_{\mu} {u^*}^{-m} \, P_{-m}^{(j)}
\lab{pmbjinv2}
\er
and
\begin{equation}
\pa^{\mu} {\tilde B}_{\mu}^{(j,\pm),{\rm inv}} + \lb A^{\mu}_{\rm inv} \, , \,
{\tilde B}_{\mu}^{(j,\pm),{\rm inv}} \rb = 0
\end{equation}
as a consequence of \rf{secondsub}.

Therefore, by using the above four types of
representations of the integrability conditions \rf{wayout}, one obtains an
infinite number of conserved currents from \rf{invcur}. They are of the form
\begin{equation}
j^{(n)}_{\mu} = {{u^*}^n \pa_{\mu} u \o {\( 1+\mid u \mid^2\)^2}}
\end{equation}
and their complex conjugates, with $n$ being any integer (positive or not).

\subsubsection{The construction of solutions for the $CP^1$ model}
\label{sec:solcp1}

As we have said in section \ref{sec:secondtype}, the gauge transformations
\rf{gauge} and \rf{newgauge} are important in the construction of the
solutions. Let us now discuss how to use them for that.  Suppose one knows a
solution $u^{(0)}$ of
$CP^1$ corresponding to the potentials $A^{(0)}_{\mu}$ and $B^{(0)}_{\mu\nu}$.
Then choosing a pair $(g,\a_{\mu})$ one obtains from \rf{gauge21}, that
\br
A_{\mu} &=&
g \, A_{\mu}^{(0)} \, g^{-1} - \pa_{\mu} g \, g^{-1} \nonu\\
B_{\mu\nu} &=&  g \, B_{\mu\nu}^{(0)} \, g^{-1}
+ g\( D_{\mu}^{A^{(0)}} \a_{\nu} - D_{\nu}^{A^{(0)}} \a_{\mu}\) g^{-1}
\lab{dressing}
\er
is also a solution of $CP^1$. What one has to do is to solve the above
equation for $u$.

As an example, consider the trivial solution of $CP^1$
\begin{equation}
u^{(0)} = {\rm constant}
\lab{vacuum}
\end{equation}
which correspond to (see \rf{cp1pota} and \rf{cp1potb})
\begin{equation}
 A_{\mu}^{(0)}=0 \,\; ; \qquad  B_{\mu\nu}^{(0)}=0
\lab{vacuumab}
\end{equation}

Then from \rf{dressing}, \rf{vacuumab}, \rf{cp1pota} and \rf{cp1potb} one
obtains that
\br
-i\pa_{\mu}g \, g^{-1} &=&
{1\o{\( 1+\mid u \mid^2\) }} \bigg(  \(\pa_{\mu} u + \pa_{\mu} u^* \)\, T_{1}
+i \(\pa_{\mu} u - \pa_{\mu} u^* \)\, T_{2}   \nonu\\
 &+&
i\( u \pa_{\mu} u^* - u^* \pa_{\mu} u \) \, T_3 \bigg)
\lab{cp1potasol}\\
-\sqrt{2}\,\epsilon_{\mu\nu\rho} g\, \pa^{\nu} \a^{\rho}  g^{-1}   &=&
{1\o{\( 1+\mid u \mid^2\)}}
 \bigg( \(\pa_{\mu} u + \pa_{\mu} u^*\)\, {\cal P}_{1}^{(1)} \nonu\\
 &+ &
i\(\pa_{\mu} u - \pa_{\mu} u^*\)\, {\cal P}_{2}^{(1)} \bigg)
\lab{cp1potbsol}
\er
where we have denoted $T_{\pm} = T_1 \pm i T_2 \;$,
$ P_{\pm 1}^{(1)} = \mp \( {\cal P}_{1}^{(1)} \pm  i {\cal P}_{2}^{(1)}\) 
/\sqrt{2}$, and
shall also denote $ P_{0}^{(1)} ={\cal P}_{3}^{(1)}$.

In order to proceed, we have to explore the representation theory of the
algebras of the type \rf{rt}. In the cases where the $P_i$'s transform under
the adjoint representation it is simple to obtain representations for \rf{rt}.
Indeed, consider
\br
\lb T_a \, , \, T_b \rb &=& f_{ab}^c T_c \nonu\\
\lb T_a \, , \, P_b \rb &=& f_{ab}^c P_c \nonu\\
\lb P_a \, , \, P_b \rb &=& 0
\lab{rtadj}
\er
If $R$ is a matrix representation for the algebra of $T_a$'s, i.e.
\begin{equation}
\lb R\( T_a \) \, , \, R\( T_b \)\rb = R\( \lb T_a \, , \, T_b \rb\)
\end{equation}
then
\br
{\hat R}\( T_a\) \equiv \(
\begin{array}{cc}
 R\( T_a \) & 0\\
0 & R\( T_a \)
\end{array} \) \; ; \qquad \qquad
{\hat R}\( P_a\) \equiv \(
\begin{array}{cc}
0 & R\( T_a \) \\
0 & 0
\end{array} \)
\lab{rhatrep}
\er
is a matrix representation of \rf{rtadj}. In addition,
\br
\Tr \( {\hat R}\( T_a\){\hat R}\( T_b\)^{\dagger} \) &=&
2 \Tr \( R\( T_a \)R\( T_b \)^{\dagger}\)
\nonu\\
\Tr \( {\hat R}\( T_a\){\hat R}\(  P_b\)^{\dagger} \) &=& 0 \nonu\\
\Tr \( {\hat R}\( P_a\){\hat R}\( P_b\)^{\dagger} \) &=&
\Tr \( R\( T_a \)R\( T_b \)^{\dagger}\)
\lab{tracestp}
\er

For the case of the $CP^1$ we have the $P$'s transforming under the adjoint
representation of $sl(2)$. In fact, $\{T_1 \, ,\, T_2\, ,\, T_3\}$ and
$\{ {\cal P}_1^{(1)}\, , \, {\cal P}_2^{(1)}\,  ,\,{\cal P}_3^{(1)}\}$,
constitute a basis satisfying \rf{rtadj} (see \rf{sl2alg}-\rf{sl2pc}).

We consider the spinor representation of $sl(2)$
\br
R\( T_1\) = \h \(
\begin{array}{rr}
0 & 1 \\
1 & 0
\end{array} \) \; ; \quad
R\( T_2\) = \h \(
\begin{array}{rrr}
0 & -i \\
i & 0
\end{array} \) \; ; \quad
R\( T_3\) = \h \(
\begin{array}{rr}
1 & 0 \\
0 & -1
\end{array} \)
\er
Using \rf{rhatrep} we get a four dimensional reducible  representation for the
algebra of $T$'s and $P$'s. We choose the basis of the representation vector
space as ($m=\h,-\h$)
\br
\mid m,(1)\rangle \equiv \(
\begin{array}{c}
\mid m\rangle \\
0
\end{array} \)  ; \quad
\mid m,(2)\rangle \equiv \(
\begin{array}{c}
0 \\
\mid m\rangle
\end{array} \) 
\er 
with 
\br
\mid 1/2\rangle \equiv \(
\begin{array}{c}
1 \\
0
\end{array} \)  ; \quad
\mid -1/2\rangle \equiv \(
\begin{array}{c}
0 \\
1
\end{array} \)
\er

Coming back to the equations \rf{cp1potasol} and \rf{cp1potbsol}, we point out
that in order to solve them, one can not consider $g$ and $\a_{\mu}$ as
independent parameters. The reason, is that the gauge potential \rf{cp1pota}
and \rf{cp1potb} are written in a given gauge, and a generic transformation
like
\rf{dressing} may take them out of that particular gauge. In the case of the
dressing method in $1+1$ dimensions \ct{dress}, the grading structure of the
Lax potentials and the corresponding generalized Gauss decomposition of the
group elements, guarantee that the transformed Lax operators are in the
original gauge. Here, we have to impose some other conditions, since we have no
gradations of the algebra playing a role. The relation between $g$ and
$\a_{\mu}$ is obtained by imposing compatibility conditions between
\rf{cp1potasol} and \rf{cp1potbsol}. One then gets
\br
i \Tr\( {\hat R}\( \pa_{\mu}g \, g^{-1}\) {\hat R}\( T_1\)\) &=&
2\sqrt{2} \epsilon_{\mu\nu\rho}\Tr\( {\hat R}\( g \pa^{\nu} \a^{\rho}
 g^{-1} \) {\hat R}\( {\cal P}_1^{(1)}\)^{\dagger}\)
\nonu\\
i \Tr\( {\hat R}\( \pa_{\mu}g \, g^{-1}\) {\hat R}\( T_2\)\) &=&
2\sqrt{2} \epsilon_{\mu\nu\rho}\Tr\( {\hat R}\( g \pa^{\nu} \a^{\rho}
 g^{-1} \) {\hat R}\( {\cal P}_2^{(1)}\)^{\dagger}\)
\nonu\\
i \Tr\( {\hat R}\( \pa_{\mu}g \, g^{-1}\) {\hat R}\( T_3\)\) &=&
{-i\o{\( 1+\mid u \mid^2\) }} \( u \pa_{\mu} u^* - u^* \pa_{\mu} u \)
\lab{niceduality1}\\
0 &=&
2\sqrt{2} \epsilon_{\mu\nu\rho}\Tr\( {\hat R}\( g \pa^{\nu} \a^{\rho}
 g^{-1} \) {\hat R}\( {\cal P}_3^{(1)}\)^{\dagger}\)
\nonu
\er

Denoting by $d\( g \)$ the adjoint representation of the group ($g T_a g^{-1} =
T^b d_{ba}\( g\)$), we have
\begin{equation}
R\( g^{-1} T_i g \) = R\( T^j \) d_{ji}\( g^{-1}\)
\end{equation}
We then have that
\br
{\hat R}\( g^{-1}\) {\hat R}\( T_i\) {\hat R}\( g\) &=& \(
\begin{array}{cc}
R\( g^{-1} T_i g \) & 0\\
0 & R\( g^{-1} T_i g \)
\end{array} \) = {\hat R}\( T_j\)\,  d_{ji}\( g^{-1}\)
\nonu\\
{\hat R}\( g^{-1}\) {\hat R}\( {\cal P}_i^{(1)}\)^{\dagger}{\hat R}\( g\)
&=& \(
\begin{array}{cc}
0 & 0\\
R\( g^{-1} T_i g \) & 0
\end{array} \) = {\hat R}\( {\cal P}_j^{(1)}\)^{\dagger}\, d_{ji}\( g^{-1}\)
\lab{adjointrhat}
\er

Therefore, one gets, from \rf{niceduality1} and \rf{adjointrhat}, that 
($d_{3j} \( g\) = \Tr \( {\hat R}\( T_j g^{-1} T_3 g\) \)$)
\br
 \Tr\( {\hat R}\(  g^{-1}\pa_{\mu}g \) {\hat R}\( T_j\)\) &-&
 \Tr\( {\hat R}\( \pa_{\mu}g \, g^{-1}\) {\hat R}\( T_3\)\) \, d_{3j}\( g\)
= \nonu\\
& & -i2\sqrt{2} \epsilon_{\mu\nu\rho}\Tr\( {\hat R}\(  \pa^{\nu} \a^{\rho} \)
{\hat R}\( {\cal P}_j^{(1)}\)^{\dagger}\)
\lab{niceduality2}
\er

Notice that \rf{niceduality2} is invariant under the local transformations
\begin{equation}
g \ra e^{i \theta T_3}\, g
\lab{localt3}
\end{equation}
The equations \rf{niceduality1} on the other hand are not invariant under 
\rf{localt3}. In fact, the set of equations \rf{niceduality1} and 
\rf{niceduality2} are not completely equivalent. 

The relation \rf{niceduality2} can be written in a more suggestive way. 
Let us denote
\begin{equation}
\a_{\mu} = \a_{\mu}^j {\cal P}_j^{(1)} \; ; \qquad \qquad
f^{\mu\nu} \equiv \pa^{\mu} \a^{\nu} - \pa^{\nu} \a^{\mu} =
f^{\mu\nu}_j {\cal P}_j^{(1)}
\end{equation}
and
\begin{equation}
{\cal J}_{\mu}^j = i \sqrt{2} \, \(
\Tr\( {\hat R}\(  g^{-1}\pa_{\mu}g \) {\hat R}\( T_j\)\) -
 \Tr\( {\hat R}\( \pa_{\mu}g \, g^{-1}\) {\hat R}\( T_3\)\) \, d_{3j}\( g\) \)
\end{equation}
Then, \rf{niceduality2} becomes
\begin{equation}
{\cal J}_{\mu}^j = \epsilon_{\mu\nu\rho} f^{\nu\rho}_j
\end{equation}

Consequently
\begin{equation}
\pa^{\mu}  {\cal J}_{\mu}^j =0
\end{equation}
and
\begin{equation}
\pa_{\nu} f^{\nu\mu}_j =  {\tilde \jmath}^{\mu}_j  \qquad\mbox{where} \qquad
{\tilde \jmath}^{\mu}_j \equiv -\h
\epsilon^{\mu\nu\rho} \pa_{\nu} {\cal J}_{\rho}^j
\end{equation}
The three conserved currents ${\cal J}_{\mu}^j$, $j=1,2,3$, must correspond
to the  currents \rf{conscurp}. Therefore, the relation \rf{niceduality2} says
that the conserved current ${\cal J}_{\mu}^j$ is dual to a field tensor 
$f^{\nu\mu}_j$. The source of that tensor in its turn is a topological current 
${\tilde \jmath}^{\mu}_j$.  
  
Notice that two  solutions for \rf{cp1potasol} are given by \rf{w1w2}.  
In fact, in the spinor representation, one finds that $W_1$ and $W_2$ are
equal~\footnote{We have checked that they are also equal in the triplet 
representation.}.
Indeed, 
\br
R(W_1) = R(W_2) = {1\over {\sqrt{1 + \mid u\mid^2}}}\, \( 
\begin{array}{cc}
1 & i u \\
i u^* & 1
\end{array} \)
\lab{gaugefixedg}
\er
and so
\begin{equation}
R(W_1)^{\dagger} = R(W_1)^{-1}
\end{equation}
Of course, the same is true for the representation ${\hat R}$ \rf{rhatrep}. 
Therefore, the $g$'s fitting into \rf{cp1potasol}  should be of the form 
\begin{equation}
{\hat R}( g ) = {\hat R} \( e^{iu^*T_-}\, e^{-\varphi T_3}\, e^{iuT_+}\)  \, 
{\hat R}(h^{-1}) = 
{\hat R} \( e^{iuT_+}\, e^{\varphi T_3}\, e^{iu^*T_-}\)  \, 
{\hat R}(h^{-1})
\lab{gwrel}
\end{equation}
where $h$ is an arbitrary constant group element.

Therefore we have that ${\hat R}(gh)={\hat R}(W_1)$, and  elements of 
such form satisfy
\begin{equation}
{\langle -1/2,(1) \mid \, \( g h\)^{\dagger} \, \mid 1/2,(1) \rangle \o
{\langle 1/2,(1) \mid \, \( g h\)^{\dagger}\, \mid 1/2,(1) \rangle}} =
- {\langle -1/2,(1) \mid \, g h \, \mid 1/2,(1) \rangle \o
{\langle 1/2,(1) \mid \, g h\, \mid 1/2,(1) \rangle}}
\lab{hermicond}
\end{equation}

Notice that not every pair $\(g , \a_{\mu}\)$ satisfying \rf{niceduality2} is a
solution for our dressing like method. The reason is that \rf{niceduality1} 
and \rf{niceduality2} are not really equivalent, and therefore a given $g$ 
satisfying  \rf{niceduality2} may not be of the form \rf{gwrel}, and so may 
not not fit into \rf{cp1potasol}.

Therefore, in order to construct the solutions we should proceed as follows. 
Choose the parameters
$\a_{\mu}$ and use \rf{niceduality2} to determine $g$. 
Notice that $\a_{\mu}$ and $\a_{\mu} + \pa_{\mu} \b$ give the same $g$.
Then choose a constant group element 
$h$ such that the symmetry \rf{localt3} may be gauge fixed in such way that 
$g$ has the form \rf{gwrel}, or equivalently ${\hat R}(gh)$ has the form 
\rf{gaugefixedg}. 
The solution for the fields $u$ and
$u^*$ are then obtained from 
\begin{equation}
u = -i {\langle 1/2,(1) \mid \, g h \, \mid -1/2,(1) \rangle \o
{\langle 1/2,(1) \mid \, g h\, \mid 1/2,(1) \rangle}}
\; ; \qquad
u^* = -i {\langle -1/2,(1) \mid \, g h \, \mid 1/2,(1) \rangle \o
{\langle 1/2,(1) \mid \, g h\, \mid 1/2,(1) \rangle}}
\lab{nicesolution}
\end{equation}
which are consequence of \rf{gwrel}. Notice that \rf{hermicond} guarantees 
the compatibility of these two relations. 

Therefore, the solutions of the $CP^1$ model, in the orbit of the vacuum
solution \rf{vacuum}, are parametrized by three abelian field tensors 
$f^{\mu\nu}_j$, $j=1,2,3$ and a constant group element $h$.
We have checked that the
baby-Skyrmion solution \rf{bsusol} is in the orbit of the vacuum \rf{vacuum},
and therefore is among the solutions discussed above.
Notice that the
matrix elements of $gh$ play the same role as the $\tau$-functions in
integrable hierarchies in $1+1$ dimensions \ct{nos}.

\subsection{The example of self-dual Yang-Mills}
\label{sec:sdym}

We consider a gauge theory, with gauge potentials $a_{\mu}$, in four 
dimensional Euclidean space-time. We impose the self-duality 
conditions
\begin{equation}
f_{\mu\nu} = {\tilde f}_{\mu\nu} \; ; \qquad \qquad 
{\tilde f}_{\mu\nu} \equiv \h \epsilon_{\mu\nu\rho\sigma}\,f_{\rho\s}
\lab{sdym1}
\end{equation}
with
\begin{equation}
f_{\mu\nu} \equiv \pa_{\mu}\, a_{\nu} - \pa_{\nu}\, a_{\mu} + 
\lb  a_{\mu} \, , \,  a_{\nu} \rb 
\end{equation}

Following Yang \ct{yang}, we now introduce the variables
\br
y &\equiv& {1\o \sqrt{2}} \( x_1 + i x_2 \) \, \qquad  
\by \equiv {1\o \sqrt{2}} \( x_1 - i x_2 \) \, \qquad \nonu\\
z &\equiv& {1\o \sqrt{2}} \( x_3 - i x_4 \) \, \qquad  
\bz \equiv {1\o \sqrt{2}} \( x_3 + i x_4 \) 
\lab{newvar} 
\er
and so, the metric is
\begin{equation}
v_{\mu}w_{\mu} = v_{y}w_{\by} + v_{\by}w_{y} +  v_{z}w_{\bz} + 
v_{\bz}w_{z}
\end{equation}

Then, the  self-duality conditions \rf{sdym1} becomes 
\br
f_{yz} &=& 0 
\lab{sdyma}\\
f_{\by\bz} &=& 0 
\lab{sdymb}\\
f_{y\by} + f_{z\bz} &=& 0
\lab{sdymc}
\er

The first two conditions, \rf{sdyma} and \rf{sdymb}, can easily be 
satisfied by taking
\br
a_{y} &=& - \pa_{y} g_1 g_1^{-1}\, , \qquad a_{z} = - \pa_{z} g_1 
g_1^{-1} \\
a_{\by} &=& - \pa_{\by} g_2 g_2^{-1} \, , \qquad a_{\bz} = 
 - \pa_{\bz} g_2 g_2^{-1} 
\lab{g1g2def}
\er
where $g_1$ and $g_2$ are two independent group elements. 

One can easily check that
\begin{equation}
f_{y\by} =  g_2 \pa_{\by} \(  \pa_{y} \cw \cw^{-1} \) g_2^{-1}
\end{equation}
where
\begin{equation}
\cw \equiv g_2^{-1} g_1 
\lab{cwdef}
\end{equation}
with a similar relation being true for $f_{z\bz}$.  

Therefore, the relation \rf{sdymc} can be written as
\begin{equation}
\pa_{\by} \(  \pa_{y} \cw \cw^{-1} \) + \pa_{\bz} \( \pa_{z} \cw 
\cw^{-1} \) = 0
\lab{goodsdym}  
\end{equation}
So, the problem of solving self-dual Yang-Mills boils down to solving 
\rf{goodsdym} \ct{yang}. That fact has been used to explore some 
properties of such system \ct{prasad,leznov,dev}. 

We now show that our integrability conditions contain the self-dual 
Yang-Mills 
as a particular case. We consider the local integrability conditions 
either of 
type {\bf I} or {\bf III}, but  with $A_{\mu}=0$ and $B_{\mu\nu}=0$. 
Therefore, 
we have the equations
\br
\cd_{\mu} {\tilde H}^{\mu} = 0 \; \qquad  \qquad 
\flie_{\mu\nu} \equiv \pa_{\mu} \ca_{\nu} - \pa_{\nu} \ca_{\mu} +
\sbr{\ca_{\mu}}{\ca_{\nu}} = 0 
\lab{4dint} 
\er
with 
\begin{equation}
\ca_{\mu} = \ca_{\mu}^r \, \ctt_r 
 \; ; \qquad  \qquad {\tilde H}_{\mu} =  {\tilde H}_{\mu}^r \, \cs_r 
\lab{4dint3} 
\end{equation}
We take $\ctt_r$ and $\cs_r$ as in \rf{rtcal}, but with the $\cs_r$'s 
transforming under the adjoint representation, i.e. 
\br
\lb \ctt_r \, , \, \ctt_s \rb &=& C_{rs}^u \ctt_u \nonu\\ 
\lb \ctt_r \, , \, \cs_s \rb &=& C_{rs}^u  \cs_u  \nonu\\
\lb \cs_r \, , \, \cs_s \rb &=& 0 
\lab{4dint4} 
\er

In terms of the new variables \rf{newvar}, 
the condition  \rf{4dint} is written as (Euclidean space-time) 
\begin{equation}
 \pa_{y} \thh_{\by} + \pa_{\by} \thh_{y} + \pa_{z} \thh_{\bz} + 
\pa_{\bz} \thh_{z} + 
\lb \ca_{y} \, , \, \thh_{\by} \rb + \lb \ca_{\by} \, , \, \thh_{y} 
\rb + 
\lb \ca_{z} \, , \, \thh_{\bz} \rb + \lb \ca_{\bz} \, , \, \thh_{z} 
\rb = 0
\lab{4dint2}
\end{equation}

We choose the potential $\ca_{\mu}$ in \rf{4dint2} to be
\begin{equation}
\ca_{\mu} = - \pa_{\mu} \cw \cw^{-1}  \equiv \ca_{\mu}^r \ctt_r
\lab{flatcala}
\end{equation}
Therefore, it is a pure gauge, and the flatness condition 
$\flie_{\mu\nu}=0$, 
in \rf{4dint}, is trivially satisfied. In addition, as a consequence 
of that,
 one has
\begin{equation}
\pa_{\mu} \ca_{\nu}^u - \pa_{\nu} \ca_{\mu}^u + 
C_{rs}^u  \ca_{\mu}^r \ca_{\nu}^s = 0
\lab{zccomp}
\end{equation}

We now  choose
\begin{equation}
\thh_{y} = \thh_{z} = 0 
\lab{currsdym1}
\end{equation}
and
\br
\thh_{\by} &=& \ca_{\by}^r \cs_r = 
-\eta^{rs} \Tr\( \pa_{\by} \cw \cw^{-1} \ctt_r\) \, \cs_s \nonu\\ 
\thh_{\bz} &=& \ca_{\bz}^r \cs_r = 
-\eta^{rs} \Tr\( \pa_{\bz} \cw \cw^{-1} \ctt_r\) \, \cs_s
\lab{currsdym2}
\er
where $\eta^{rs}$ is the inverse of $\eta_{rs} \equiv \Tr \( \ctt_r 
\ctt_s \)$. 

Then, substituting in \rf{4dint2} and using \rf{zccomp} one gets
\begin{equation}
\pa_{\by} \ca_{y}^r + \pa_{\bz} \ca_{z}^r = 0
\end{equation}
Contracting with $\ctt_r$ one gets the condition of self-duality 
\rf{goodsdym}.

Therefore, from \rf{currsdym1} and \rf{currsdym2}, one gets that the 
conserved 
quantities \rf{conscurcal} are given by
\begin{equation}
{\cal J}_z = {\cal J}_y =0 \; ; \qquad 
{\cal J}_{\by} = \ca_{\by}^r  \, \cw^{-1} \cs_r \cw \; ; \qquad 
{\cal J}_{\bz} = \ca_{\bz}^r  \, \cw^{-1} \cs_r \cw 
\end{equation}
and the conservation law is
\begin{equation}
\pa_{\mu} {\cal J}^{\mu} = 0 \qquad  \ra \qquad 
\pa_{y}{\cal J}_{\by} + \pa_{z}{\cal J}_{\bz} = 0
\lab{conslawsdym}
\end{equation}

Since the $\cs_r$'s and $\ctt_r$'s transform under the adjoint 
representation,  
 ($d(\cw_1)d(\cw_2)=d(\cw_1\cw_2)$)
\begin{equation}
\cw^{-1} \cs_r \cw = \cs^s \, d_{sr}(\cw^{-1}) \; ; \qquad \qquad 
\cw^{-1} \ctt_r \cw = \ctt^s \, d_{sr}(\cw^{-1}) 
\end{equation}
one gets that \rf{conslawsdym} is the same as  
\begin{equation}
\pa_{y}\( \cw^{-1}\pa_{\by}\cw \) + \pa_{z}\( \cw^{-1} \pa_{\bz}\cw 
\) = 0
\end{equation}
But that is equivalent to \rf{goodsdym}. So, the self-duality 
equations  
\rf{goodsdym} are themselves the conservation laws.

\subsection{The example of the Bogomolny equations}

We consider a gauge theory, with gauge potentials $a_{\mu}$, with Higgs field
$\vp$ in the adjoint representation, on a Minkowski space-time with
coordinates $x^{\mu}$, $\mu = 0,1,2,3$. The equations of motion are 
\br
\nabla_{\mu} f^{\mu\nu} &=& \lb \vp \, , \, \nabla^{\nu}\vp \rb \nonu\\
\nabla_{\mu} {\tilde f}^{\mu\nu} &=& 0 \nonu\\
\nabla_{\mu}\nabla^{\mu} \vp_r &=& - {\pa V\( \vp\) \o \pa \vp_r} 
\lab{higgsmodel}
\er
where $\nabla_{\mu} * \equiv \pa_{\mu} \, * + \lb a_{\mu} \, , \, * \rb$, 
${\tilde f}^{\mu\nu} \equiv \h \epsilon^{\mu\nu\rho\s} f_{\rho\s}$, and 
$\vp = \vp_r \ctt^r$. 

We shall consider the so-called Prasad-Sommerfield limit \ct{ps} where the
Higgs potential is taken to vanish, but the vacuum expectation value of  the
Higgs field is kept different from zero. We then consider the
solutions of \rf{higgsmodel} satisfying 
\begin{equation}
f^{0i} = 0 \; ; \qquad \qquad \nabla_{0} \vp =0 
\lab{staticcond}
\end{equation}
and the Bogomolny equation \ct{bogo}
\begin{equation}
f_{ij} = \epsilon_{ijk} \nabla_k \vp 
\lab{bogoeq}
\end{equation}
with $i,j,k=1,2,3$. 

One can verify that the first order differential equations \rf{staticcond} and 
\rf{bogoeq} imply the second order equations \rf{higgsmodel} 
(for $V\( \vp\)=0$). The relations \rf{staticcond} are automatically satisfied
if we consider static configurations in the gauge $a_0=0$. 

Therefore what one has  to solve are  the Bogomolny equations \rf{bogoeq}.
However, those can be written as a self-duality equation as follows. Since the
Higgs field is in the adjoint representation one can introduce  a further
component for the gauge potential as
\begin{equation}
a_4\equiv \vp 
\end{equation}
In addition, one introduces an extra space-time coordinate $x_4$, such that
$x_m$, $m=1,2,3,4$, are the coordinates of a Euclidean space. However, the
fields are taken to be independent of $x_4$. Therefore,
\begin{equation}
f_{i4} = \pa_i \vp + \lb a_i \, , \, \vp \rb = \nabla_{i}\vp 
\end{equation}
and \rf{bogoeq} can be written as
\begin{equation}
f_{mn} = {\tilde f}_{mn} \, ; \qquad 
{\tilde f}_{mn} \equiv \h  \epsilon_{mnpq} f_{pq} \; ; \qquad 
m,n,p,q=1,2,3,4
\lab{sdbogo}
\end{equation} 
Therefore, we have the same situation as in the case of the self-dual
Yang-Mills discussed in section \ref{sec:sdym}. So, following \rf{g1g2def} we
write (we are in the gauge $a_0 = 0$)
\br
a_1 - i a_2 &=& - \(\pa_1 - i \pa_2 \) g_1 g_1^{-1} \, \qquad 
a_3 + i \vp = - \pa_3 g_1 g_1^{-1} \nonu\\
a_1 + i a_2 &=& - \(\pa_1 + i \pa_2 \) g_2 g_2^{-1} \, \qquad 
a_3 - i \vp = - \pa_3 g_2 g_2^{-1}
\er
Substituting these equations in \rf{sdbogo} one gets that the only non trivial
equation left is (in analogy to \rf{goodsdym})
\begin{equation}
\sum_{i=1}^3 \pa_i \( \pa_i \cw \cw^{-1} \) = 
i \lb \pa_1 \cw \cw^{-1} \, , \,\pa_2\cw \cw^{-1} \rb 
\lab{goodsdbogo}
\end{equation}
where (as in \rf{cwdef}) we have defined $\cw \equiv g_2^{-1} g_1$. 

The Bogomolny equations can then be written as our integrability conditions of
the type {\bf I} or {\bf II}, with $A_{m}=0$ and $B_{mn}=0$, following
what we have done in the case of the self-dual Yang-Mills.  
Such conditions are written however, in the Euclidean space with coordinates 
$x_m$, $m=1,2,3,4$. 
We take
${\tilde H}_{m}$ and $\ca_{m}$ satisfying \rf{4dint}, \rf{4dint3},
\rf{4dint4} and \rf{flatcala}. The relation \rf{goodsdbogo} is then the same
as the covariant divergence of ${\tilde H}_{m}$. More explicitly we have  
\begin{equation}
{\tilde H}_1 = i {\tilde H}_2 \; ; \qquad {\tilde H}_3 = - i {\tilde H}_4
\end{equation}
and
\br
\thh_{1} &=& \h \( \ca_{1}^r +i \ca_{2}^r\) \cs_r = 
-\h \eta^{rs} \Tr\( \( \pa_{1} + i \pa_{2}\)\cw \cw^{-1} \ctt_r\) \, \cs_s
\nonu\\ 
\thh_{3} &=& \h \ca_{3}^r  \cs_r = 
-\h \eta^{rs} \Tr\( \pa_{3} \cw \cw^{-1} \ctt_r\) \, \cs_s 
\lab{currsdbogo}
\er

The conserved currents are 
\begin{equation}
{\cal J}_1 = i{\cal J}_2  \; ; \quad 
{\cal J}_3 = -i{\cal J}_4  \; ; \quad 
{\cal J}_{1} = \h\( \ca_{1}^r +i\ca_{2}^r \) \, \cw^{-1} \cs_r \cw \; ; \quad 
{\cal J}_{3} = \h \ca_{3}^r  \, \cw^{-1} \cs_r \cw 
\end{equation}

\vspace{1 cm}

\section*{Acknowledgments} 

We are grateful to R.~Aldrovandi, Ch.~Devchand, D.~Gianzo, J.F.~Gomes, 
M.~Mari\~no, J.L.~Miramontes, R.~Nepomechie, D.I.~Olive, A.~Ramallo, 
M.V.~Saveliev, and A.H.~Zi\-mer\-man for helpful discussions.  OA would 
like to thank the Iberdrola Foundation which helped initiate this 
collaboration.  OA is partially supported by NSF grant PHY-9507829.  
LAF is partially supported by CNPq (Brazil).  JSG is partially 
supported by EC TMRERBFMRXCT960012.  OA and LAF would like to thank 
the University of Santiago de Compostela for its hospitality.

\end{document}